\documentclass{jfm}
\usepackage{graphicx,amssymb,amsmath}
\usepackage{natbib,lscape}
\usepackage{subfig}
\usepackage{rotating,framed,color}
\definecolor{shadecolor}{gray}{0.8}

\title{Chaotic Rotation of a Towed Elliptical Cylinder}

\author[G. D. Weymouth] {G.\ns D.\ns W\ls E\ls
  Y\ls M\ls O\ls U\ls T\ls H\ls 
  \thanks{Email address for correspondence: G.D.Weymouth@soton.ac.uk}}

\affiliation{Southampton Marine and Maritime Institute, University of Southampton, Southampton, UK \\ }

\label{firstpage}
\begin{document}

\maketitle

\begin{abstract}

In this paper I consider the self-excited rotation of an elliptical cylinder towed in a viscous fluid as a canonical model of nonlinear fluid structure interactions with possible applications in the design of sensors and energy extraction devices. First, the self-excited ellipse system is shown to be analogous to the forced bistable oscillators studied in classic chaos theory. A single variable, the distance between the pivot and the centroid, governs the system bifurcation into bi-stability. Next, fully coupled computational fluid dynamics simulations of the motion of the cylinder demonstrate limit cycle, period doubling, intermittently chaotic, and fully chaotic dynamics as the distance is further adjusted. The viscous wake behind the cylinder is presented for the limit cycle cases and new types of stable wakes are characterized for each. In contrast, a chaotic case demonstrates an independence of the wake and structural states. The rotational kinetic energy is quantified and correlated to the vortex shedding and the trajectory periodicity. Chaotic and high-period system responses are found to persist when structural damping is applied and for Reynolds numbers as low as 200.

\end{abstract}


\section{Introduction}\label{intro}

The analysis of fluid structure interactions (FSI) is fundamental to diverse engineering fields with examples ranging from the flow induced motions of telephone wires and offshore drilling risers \citep{Bearman1984,Williamson2004}, to the flutter of membranes such as the wheezing of the soft pallet \citep{Huang1995} or a flapping flag \citep{Connell2007}. Exciting new developments in the field include the design of energy extraction devices which utilize resonant structural response to the flow \citep{Bernitsas2008,Abdelkefi2013,Barrero2010} and passive hydrodynamic sensors such as those of \cite{Beem2013} which are inspired by the ability of a hunting harbour seal to detect its prey's wake minutes after its passing.

While some of these systems are approximately linear and many feature periodic responses, the non-periodic response of nonlinear deterministic systems have been a source of fascination and rewarding research since their discovery by \cite{Lorenz1963}. \cite{Holmes1979} and others developed this theory to apply to a broad range of harmonically forced model systems useful in mechanical applications. Chaotic FSI gives rise to many practical problems such as snap loading \citep{Connell2007}, buckling of wings in aircraft manoeuvres \citep{Sipcic1990} and ship roll and capsize \citep{Spyrou2000}. The broad frequency response of chaotic mechanical systems is also important, with its impact on the fatigue life of structures \citep{Modarres2011,Dahl2007}, and its ability to maximize response of sensors and energy extraction devices over a broad range of excitation frequencies \citep{Arrieta2010,Townsend2013}.

In this paper I consider the self-excited rotation of an elliptical cylinder towed through a viscous fluid as a canonical model of nonlinear fluid structure interactions. Like a simple pendulum, this is a nearly trivial one degree of freedom mechanical system, yet its behaviour is incredibly rich; sensitively ranging through periodic, semi-periodic, and fully chaotic dynamics. The investigation of simple nonlinear FSI systems is an active one and recent experimental work has shown that a single degree of freedom pendular disk in a cross flow exhibits bistability \citep{Obligado2013}. An entire field of literature exists on the rotational galloping and torsional flutter of bluff cylinders due to their prevalence in civil and industrial engineering \citep{Nakamura1990,Oudheusden1996a,Robertson2003,Alonso2010}. Understandably these analyses are mostly focused on preventing such large amplitude motions and generally feature quasi-static analysis and static experiments at high Reynolds number. In contrast, the present study is focused on describing the nonlinear dynamics of this system free from structural restoring forces. Additionally, the current study uses low Reynolds number ($Re\leq 10^3$) both because this is more appropriate for small scale flow sensors, but also because it avoids turbulent fluid forcing confusing the analysis of this deterministic system. Another related area of research is that of autorotation \citep{Lugt1983}, in which an object such as a flat plate rotates about an axis perpendicular to an oncoming flow despite having a nominally zero net torque about its center. The rotation is due to the unsteady vortex shedding, and while thick ellipses show much less tendency to autorotate than plates due to their rounded edges they are near the stability limit \citep{Lugt1980}.

The towed rotating ellipse is a uniquely simple example of chaotic FSI. While the fluid supplies infinitely many degrees of freedom and nonlinear forcing even at low Reynolds number, previous example systems depend strongly on continuously deformable or multi-linkage structures or on turbulent flow. Indeed, the onset of chaotic motions in long flexible risers is attributed to the interaction of many structural modes in the inline and crossflow directions along the span \citep{Modarres2011}. And flapping membranes with excessive stiffness or reduced mass-ratio exhibit only simple period-1 oscillations up to $Re=5000$ \citep{Connell2007}. In the one degree of freedom rotational galloping experiments of \cite{Oudheusden1996a} two stable limit cycles are found for certain parameters, but the author explains the difficulty in isolating this possible bifurcation experimentally. The current work is therefore not only beneficial because of the problem's relevance to engineering systems, but also because of its simplicity and the possibility for detailed investigation using fully coupled computational simulations. 

The formulation of the system is detailed in \S\ref{formulation}, using analytic arguments to demonstrate its similarity to the classic forced bistable oscillators studied in the early development of chaos theory. Next, \S \ref{method} presents the numerical approach used to study the fully coupled unsteady fluid-structure system. In \S \ref{results} the trajectories, wake modes, and energy levels are presented over a range of geometric, structural damping, and $Re$ conditions. Finally, \S \ref{conclusion} discusses the findings and presents conclusions.  

\section{Formulation and Equivalence to Forced Bistable Oscillators}\label{formulation}

\begin{figure}
	\centering
	\subfloat[]{
		\includegraphics[width=0.3\textwidth]{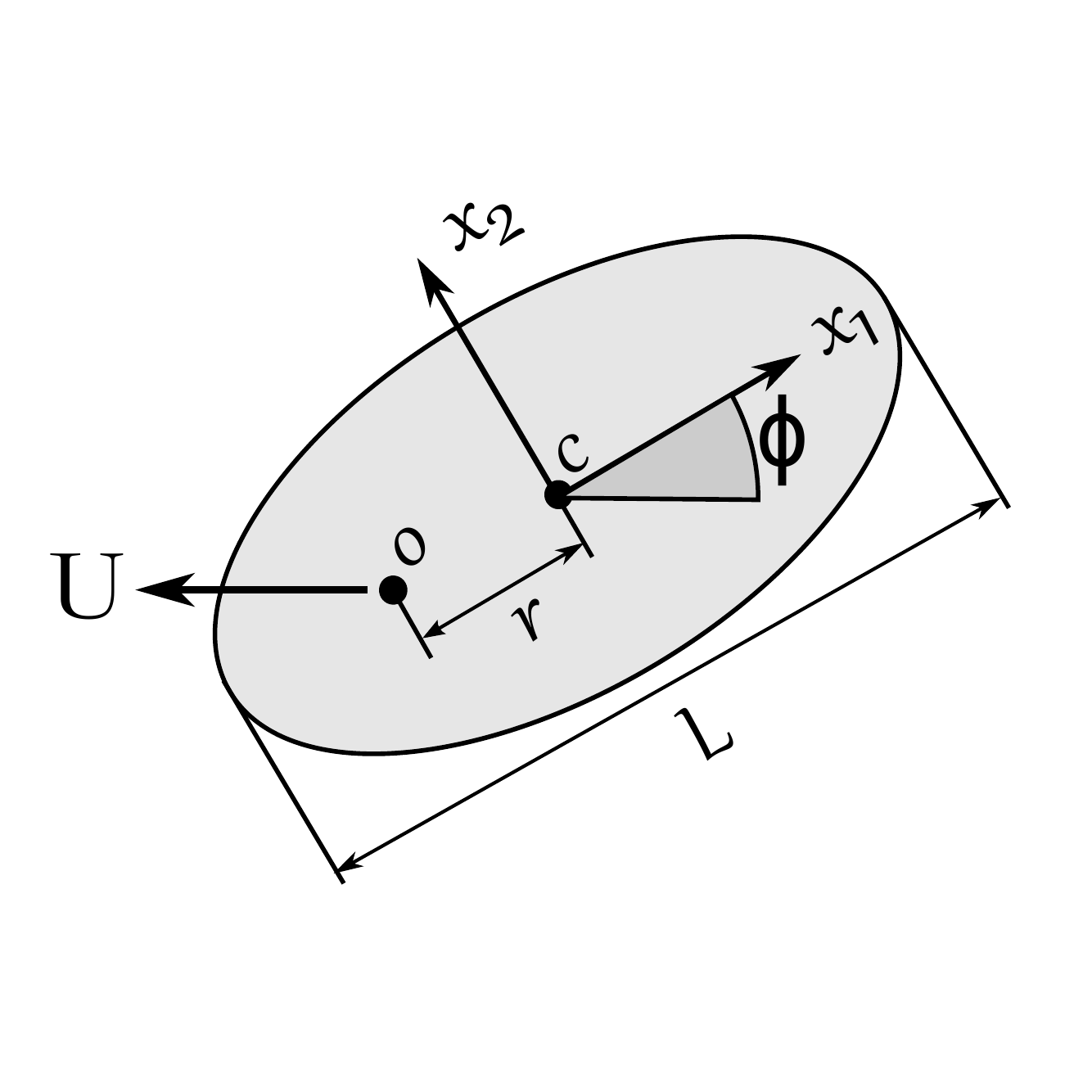}
		\label{fig: drawing}}
	\subfloat[]{
		\includegraphics[width=0.35\textwidth]{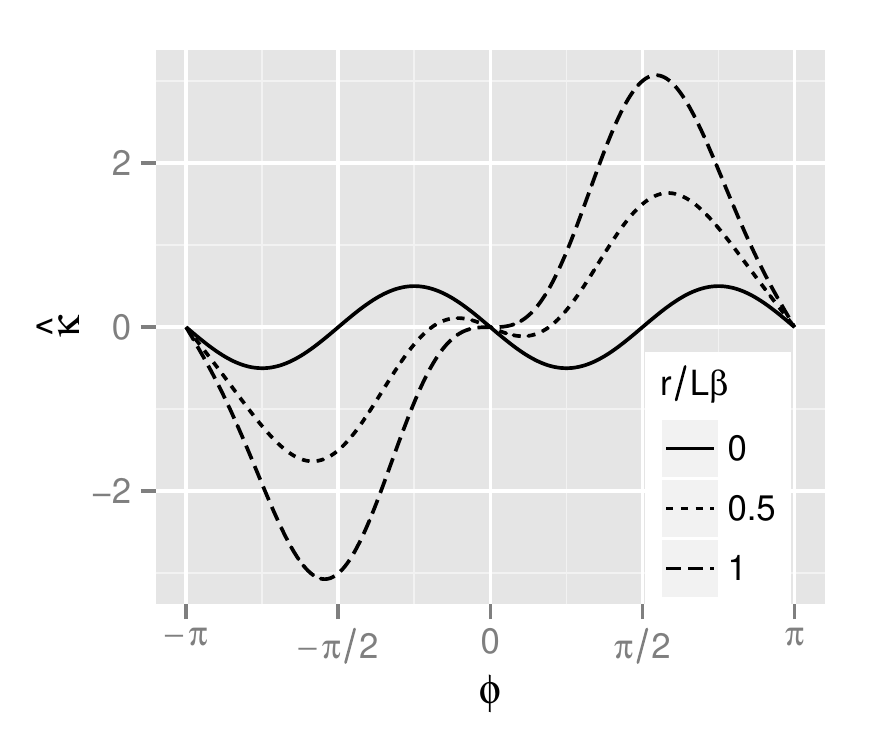}
		\label{fig: torque}}
	\subfloat[]{
		\includegraphics[width=0.35\textwidth]{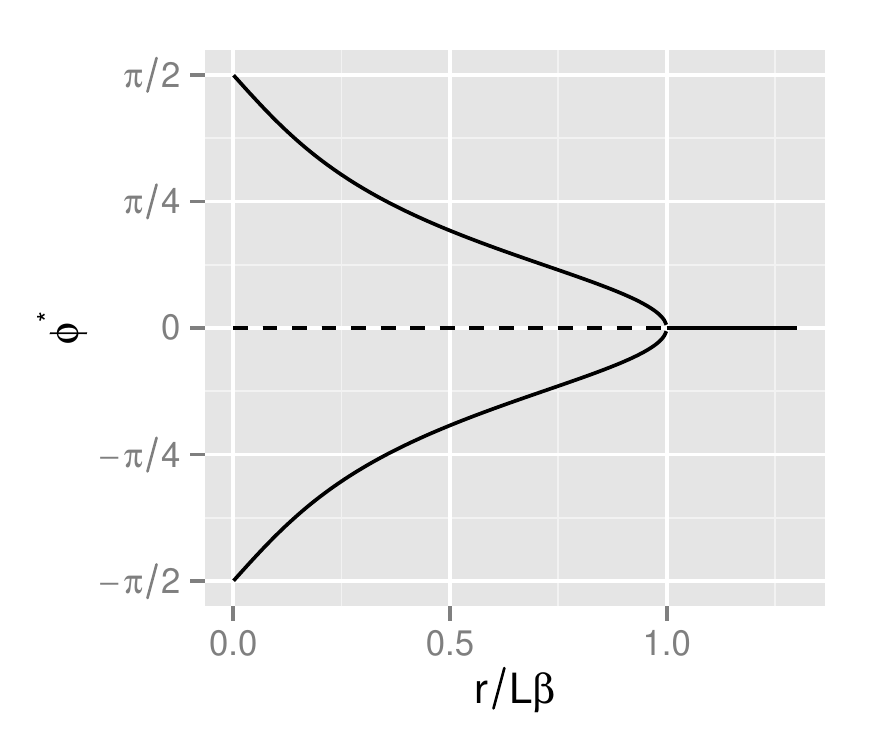}
		\label{fig: spring}}
	\caption{(a) Sketch of the fluid structure interaction problem. The ellipse is towed at constant speed and allowed to rotate freely around the pivot at  $o$. (b) Approximate normalized restoring torque $\hat\kappa$ given by equation \ref{eq: drag torque}. (c) Fixed points $\phi^*$, such that $\hat\kappa(\phi^*) = 0$, as a function of $r/L\beta$. Stable points are solid; unstable are dashed.}
\end{figure}

Consider a two-dimensional elliptical cylinder, towed at steady velocity $U$ through a fluid of density $\rho$, and free to pivot about the tow point. The system is sketched in Fig~\ref{fig: drawing}. The major axis of the ellipse is $L$, the minor axis is $l$, and the instantaneous angle of rotation is $\phi$. The pivot point is located a distance $r$ from the centroid of the ellipse along the major axis.

While the structure has only one degree of freedom, the dynamical system also depends on state of the continuous fluid. The angular equation of motion about the pivot governs the rotation of the ellipse,
\begin{equation} \label{eq: ang mom}
I_o \ddot\phi +\zeta \dot\phi = \tau_o = \oint_S \vec f(s) \times (\vec x(s) -\vec x_o)  ds
\end{equation}
where $I_o$ is the second moment of area of the ellipse about the pivot, $\ddot\phi$ is the angular acceleration, $\zeta$ is the structural damping, $\dot\phi$ is the angular velocity, and $\tau_o$ is the integrated torque relative to the pivot due to the contact force $\vec f$ on the solid/fluid interface $S$. The resulting position and velocity of the ellipse set the boundary conditions on the fluid, leading to a fully coupled FSI problem.

The fluid motion and resultant forces in Eq \ref{eq: ang mom} are governed by the full viscous Navier-Stokes equation, but it is insightful to approximate these forces analytically where possible. In particular, given that the added mass matrix for the ellipse is diagonal, the potential flow estimate is,
\begin{eqnarray} \label{eq: inviscid torque}
\tilde \tau_o &=& -(m_{66}+r^2 m_{22}) \ddot\phi+\frac 12 U^2 \sin 2\phi (m_{22}-m_{11}) \\
 &=& -I_a \ddot\phi - \kappa_M(\phi)
\end{eqnarray}
where $(m_{11},m_{22},m_{66})=\frac 14 \rho\pi\left(l^2,L^2, \frac 1 {32} (L^2-l^2)^2\right)$ are the added mass components in each body fixed coordinate direction $(x_1,x_2,\phi=x_6)$ as labelled in Fig \ref{fig: drawing}. Equation \ref{eq: inviscid torque} is composed of an angular fluid inertia term $I_a$ from the acceleration of the off-center ellipse, and the well known Munk moment $\kappa_M(\phi)$ which tends to rotate slender bodies broadside to the flow. Indeed the stable fixed points for this system are  $\phi^*_s=\pm\frac\pi2$ while the unstable fixed points are $\phi^*_u=0,\pi$. 

Note that the Munk moment has no dependence on the lever arm $r$. In thin airfoil theory the Munk moment is supplemented with the torque induced by the lift force, but this cannot be applied the flow around a bluff body. A simple drag force model,
\begin{equation} \label{eq: drag}
D = \frac 12 C_D(\phi) \rho U^2 L 
\end{equation}
is more appropriate, where the drag coefficient could be roughly estimated as $C_D = 1-\frac 12 \cos 2\phi$. Additionally, the lack of pressure recovery on the back half of the ellipse which causes the drag will also reduce the magnitude of the Munk moment \citep{Lugt1980}. Applying this drag force to the centroid of the ellipse gives the estimated total restoring torque as,
\begin{equation} \label{eq: drag torque}
\hat{\kappa} = \frac{\kappa}{\frac\beta 4 \rho U^2L^2} = -\sin \phi \left[\cos\phi- \frac r {L\beta} (2-\cos2\phi)\right]
\end{equation}
where $\beta$ is an empirical coefficient scaling the Munk moment. When the pivot is at the center ($r/L\beta=0$) the torque is given soley by the Munk moment and the stable fixed points are located at $\phi^*_s=\pm\frac\pi2$, as before. Increasing the size of the lever arm increases the drag component of the restoring torque, driving the stable fixed points toward $\phi=0$ (Fig~\ref{fig: torque},\ref{fig: spring}). At $r/L\beta=1$, the two stable points merge at $\phi=0$ and this fixed point undergoes a Hopf bifurcation and becomes the sole stable fixed point.

The arrangement of two stable fixed points straddling an unstable fixed point classifies the system as a bistable oscillator. The classic bistable model system considered in \cite{Holmes1979} is
\begin{equation}\label{eq: holmes}
\ddot \phi +\zeta \dot \phi - c_1\phi + c_2\phi^3 = f\cos(\omega t)
\end{equation}
and informative comparisons can be made to this system. Performing a Taylor expansion of Eq \ref{eq: drag torque} around $\phi=0$ recovers $\hat\kappa = -c_1\phi+c_2\phi^3+O(\phi^5)$ where $c_1 = 1- r/ L\beta$ and $c_2 = \frac {11}6 r /L\beta+\frac 23$. The system has a cubic restoring term, which leads to its bistability, as in Eq~\ref{eq: holmes}. However, the nonlinearity in the rotating ellipse system is more severe. First, the polynomial coefficients of the ellipse restoring torque are a function of $\phi$. The spring torque at the stable fixed point varies strongly with  $r/L\beta$ and has a local maximum at $r/L\beta\simeq 0.2$ which can not be modeled by Eq~\ref{eq: holmes}. Second, while the motion through a viscous fluid will supply damping, that damping is nonlinear and depends on both $\phi$ and $\dot\phi$. Finally, time-dependent forcing similar to that in Eq~\ref{eq: holmes} is supplied by the alternating shedding of vortices off the body, but this will not produce a constant amplitude harmonic force. Instead, the amplitude and timing depend on the state of the near-field flow around the body. These additional nonlinearities require the use of a numerical approach to fully quantify the response of the towed ellipse.

\section{Numerical Methodology for the Fully Coupled Simulations}\label{method}

A set of two-dimensional simulations enable detailed quantitative predictions of the coupled viscous fluid / dynamic body system. In this work I utilize the Boundary Data Immersion Method (BDIM), a robust immersed boundary method suitable for dynamic fluid-structure interaction problems detailed in \cite{Weymouth2006} and \cite{Weymouth2011JCP}. Briefly, the full Navier-Stokes equations and the angular governing equation \ref{eq: ang mom} are convolved  with a kernel of support $\epsilon=2h$, where $h$ is the grid spacing. The integrated equations are valid over the complete domain and allow for general solid body dynamics to be simulated. Previous work has validated this approach for a variety of dynamic rigid-body problems such as flapping and translating foils \citep{Wibawa2012} and  deforming body problems such as a model of a fast escaping octopus \citep{Weymouth2013JFM}.

The tests are run using a 2:1 ellipse and the body density matches the fluid density. An inertial computational domain is used with dimensions $8L \times 5L$ which translates with the body but does not rotate. All cases use the no-slip and no-penetration boundary conditions on the solid/fluid interface. No-penetration conditions are applied on the top and bottom walls and a convection exit condition is used. This narrow domain incurs blockage effects but tests with larger computational domains indicated that the same nonlinear response features are found, just at slightly different values of the pivot location. 

\begin{table}
\vspace{-.5cm}
\caption{Amplitude of oscillation $\Phi$ and percentage error relative to the finest solution $\epsilon=1-\Phi/\Phi_{h=0.01L}$ for the $r=0$  period-1 limit cycle presented in \S\ref{results}.}
\centering 
\begin{tabular}{cccccc} 
\hline\hline
$h/L$      & 0.039 & 0.026 & 0.02 & 0.013 & 0.01 \\ [-0.5ex]
\hline
$\Phi$     & 0.421 & 0.470 & 0.497 & 0.504 & 0.507 \\ 
$\epsilon$ &  0.17 & 0.074 & 0.020 & 0.007 & |\\
\hline 
\vspace{-.5cm}
\end{tabular}
\label{tab: error} 
\end{table}

The coupled BDIM equations are discretized using a finite-volume method (third-order convection and second-order diffusion) in space and Heun's explicit second-order method in time. An adaptive time-stepping scheme is used to maintain stability. Table \ref{tab: error} presents a grid convergence study on the magnitude of oscillation for a periodic case with $r=0$. The results converge with third-order accuracy overall and the maximum difference in the solution between the fine and intermediate grid ($h/L=0.02$) is only 2\%. This verifies the high accuracy of these viscous two-dimensional simulations and the intermediate grid level is used for the remainder of the paper. 

\section{Results for the Freely Rotating Ellipse}\label{results}

This section details the numerical results of the self-excited rotations of the 2:1 neutrally buoyant elliptical cylinder. The Reynolds number based on the steady tow speed $U$ is set to $Re=\frac{UL}{\nu}=10^3$ unless otherwise mentioned to avoid intrinsically non-periodic forcing due to transition to turbulent flow on the cylinder. The length of the lever arm was systematically varied in the range of $0\leq r/L \leq \frac 12 $ which simultaneously adjusts the fluid restoring torque and moment of inertia as detailed in \S\ref{formulation}. The structural damping $\zeta$ is set to zero for all but the final tests.  

The analysis in \S\ref{formulation} identified stable and unstable fixed points of the system. However, none of the cases studied with the 2:1 ellipse produced fixed point motion due to the spontaneous onset of bluff body vortex shedding. The resulting forces cause the fixed points to undergo a subcritical bifurcation, resulting in trajectories that show either limit-cycle, intermittently periodic, or fully chaotic motion.

\begin{figure}
	\centering
	\begin{minipage}{0.38\textwidth}\centering
	\begin{snugshade}
		\scalebox{0.8}{$t/T = 0 $} 
	\end{snugshade}
	\end{minipage}
	\hspace{1mm}
	\begin{minipage}{0.38\textwidth}\centering
	\begin{snugshade}
		\scalebox{0.8}{$t/T \simeq 0.5 $} 
	\end{snugshade}
	\end{minipage}
	\begin{minipage}{5mm}
	$\ $
	\end{minipage}\\
	\begin{minipage}{0.4\textwidth}
		\scalebox{1}[-1]{\includegraphics[trim=4cm 7cm 2cm 7cm, clip=true, 
		width=\textwidth]{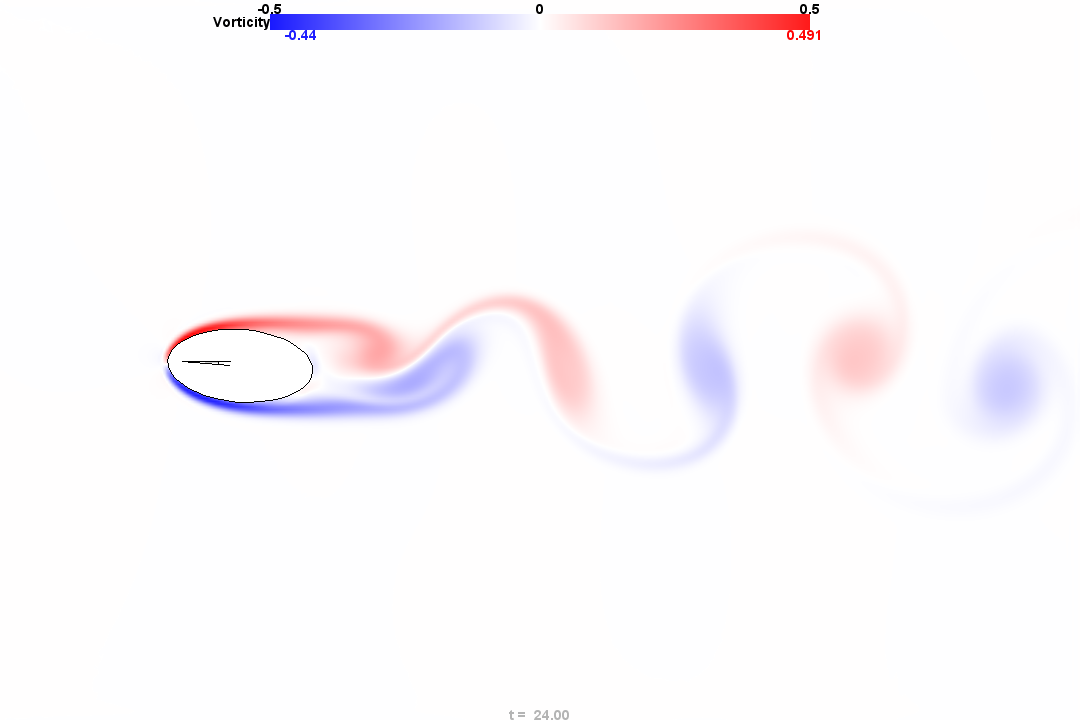}}
	\end{minipage}
	\begin{minipage}{0.4\textwidth}
		\scalebox{1}[-1]{\includegraphics[trim=4cm 7cm 2cm 7cm, clip=true, 
		width=\textwidth]{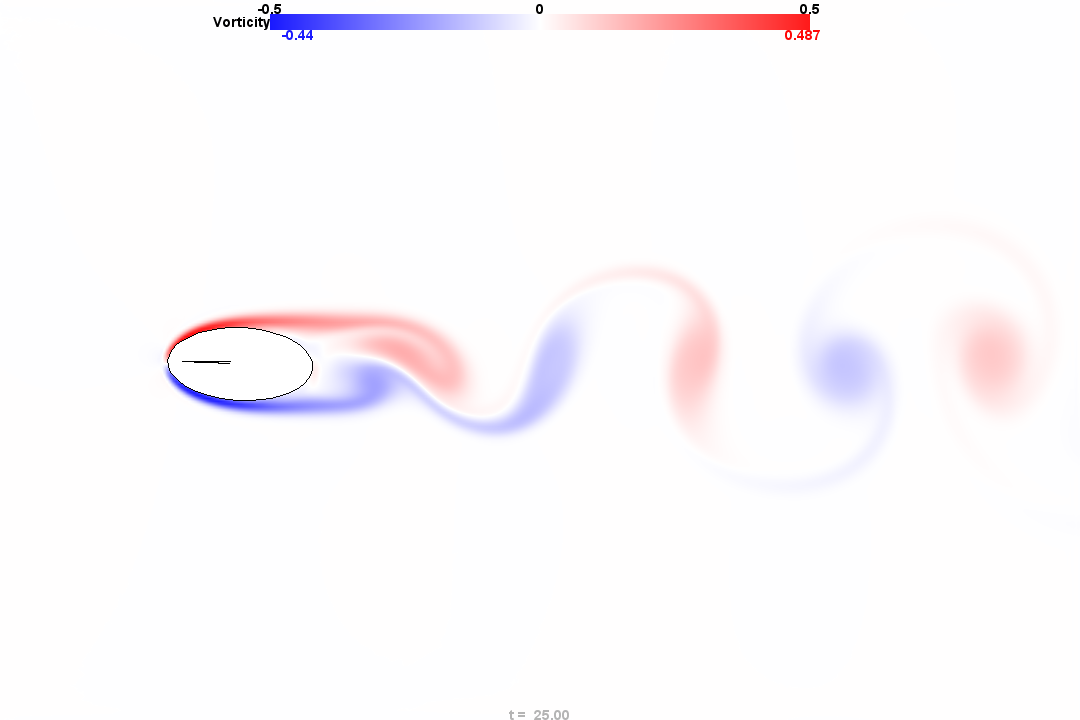}}
	\end{minipage}
	\rotatebox[origin=c]{270}{
	\begin{minipage}{16mm}\centering
	\begin{snugshade}
		\scalebox{0.8}{$r/L = 0.4 $} 
	\end{snugshade}
	\end{minipage}}\\
	\begin{minipage}{0.4\textwidth}
		\scalebox{1}[-1]{\includegraphics[trim=4cm 6cm 2cm 4cm, clip=true, 
		width=\textwidth]{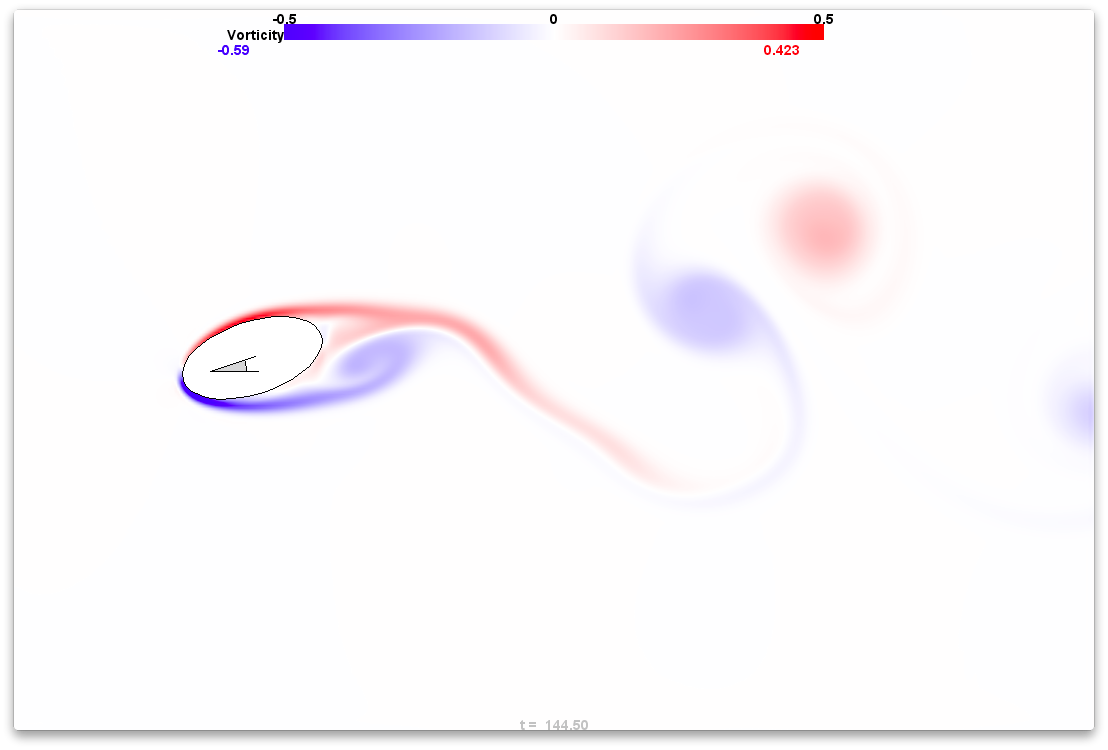}}
	\end{minipage}
	\begin{minipage}{0.4\textwidth}
		\scalebox{1}[-1]{\includegraphics[trim=4cm 6cm 2cm 4cm, clip=true, 
		width=\textwidth]{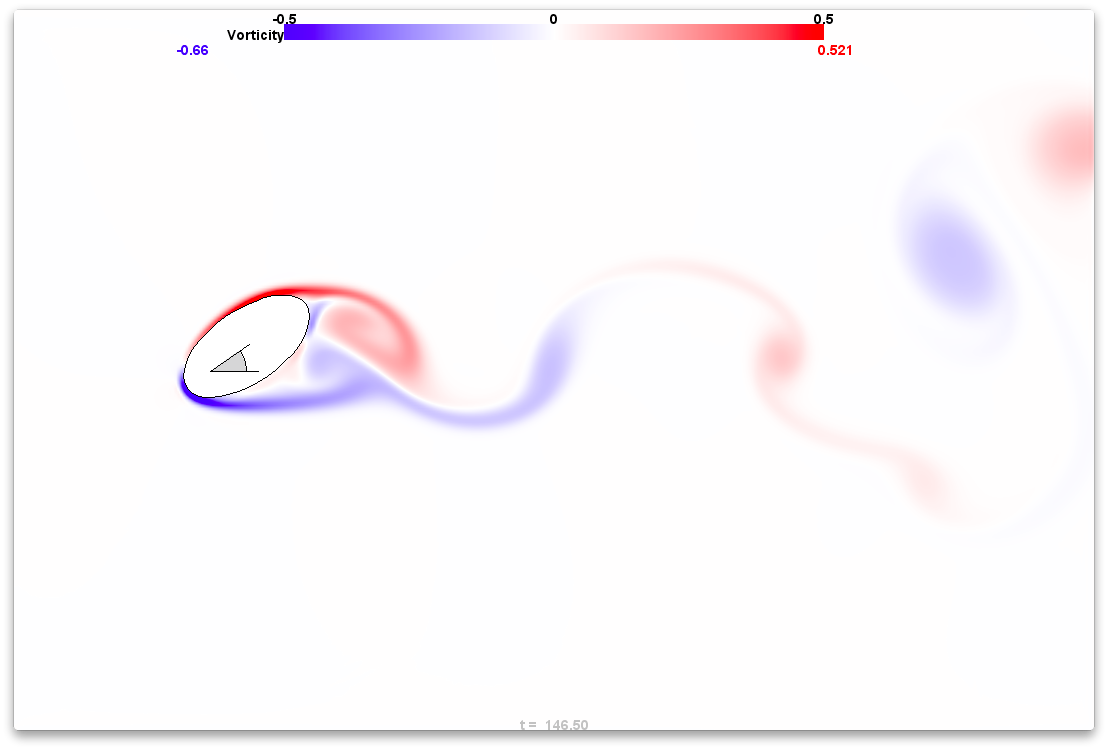}}
	\end{minipage}
	\rotatebox[origin=c]{270}{
	\begin{minipage}{23mm}\centering
	\begin{snugshade} 
		\scalebox{0.8}{$r/L = 0.3$ }
	\end{snugshade}
	\end{minipage}}\\
	\begin{minipage}{0.4\textwidth}
		\scalebox{1}[-1]{\includegraphics[trim=4cm 4cm 2cm 6cm, clip=true, 
		width=\textwidth]{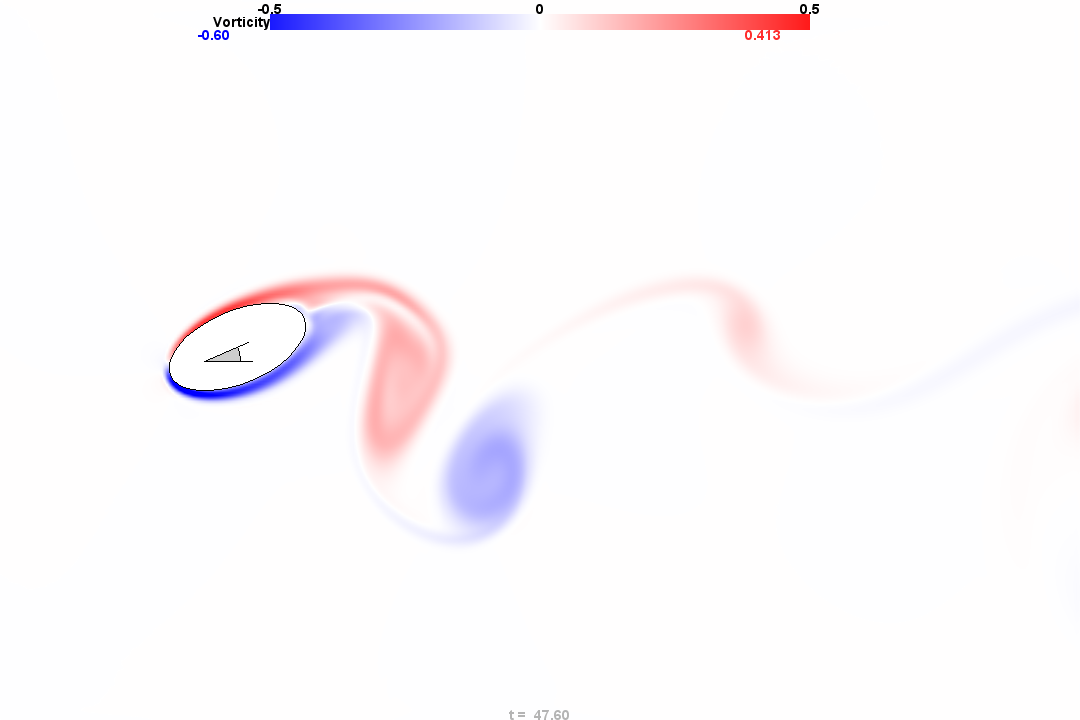}}
	\end{minipage}
	\begin{minipage}{0.4\textwidth}
		\scalebox{1}[-1]{\includegraphics[trim=4cm 4cm 2cm 6cm, clip=true, 
		width=\textwidth]{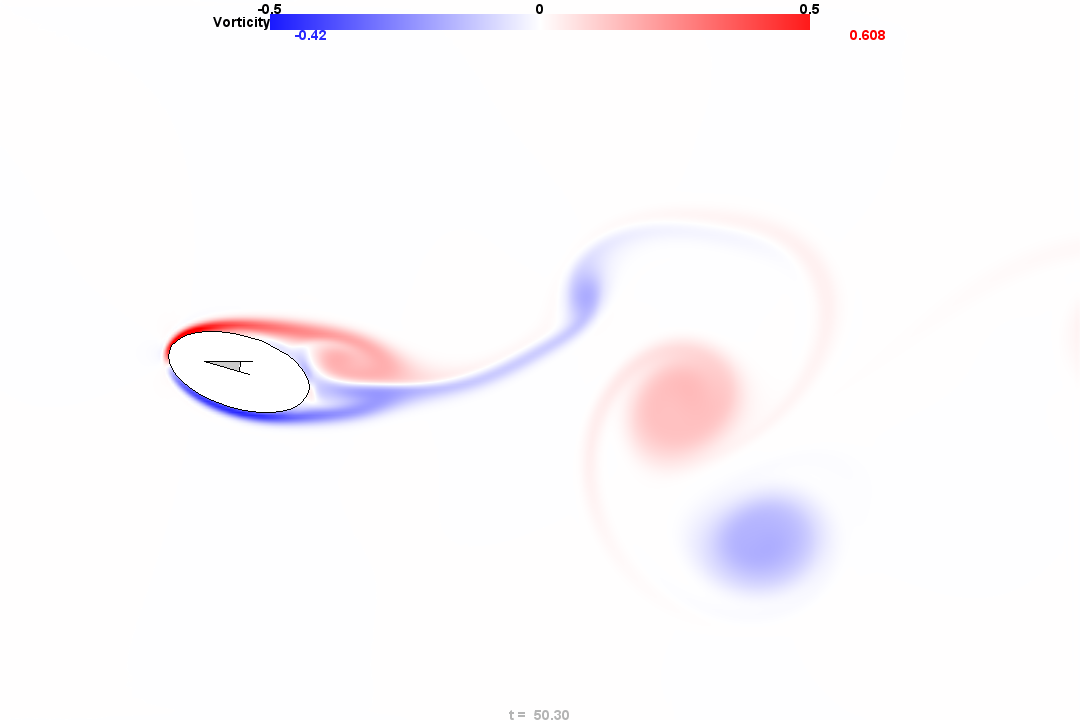}}
	\end{minipage}
	\rotatebox[origin=c]{270}{
	\begin{minipage}{23mm}\centering
	\begin{snugshade} 
		\scalebox{0.8}{$r/L = 0.25$ }
	\end{snugshade}
	\end{minipage}}\\
	\begin{minipage}{0.4\textwidth}
		\scalebox{1}[-1]{\includegraphics[trim=4cm 4cm 2cm 4cm, clip=true, 
		width=\textwidth]{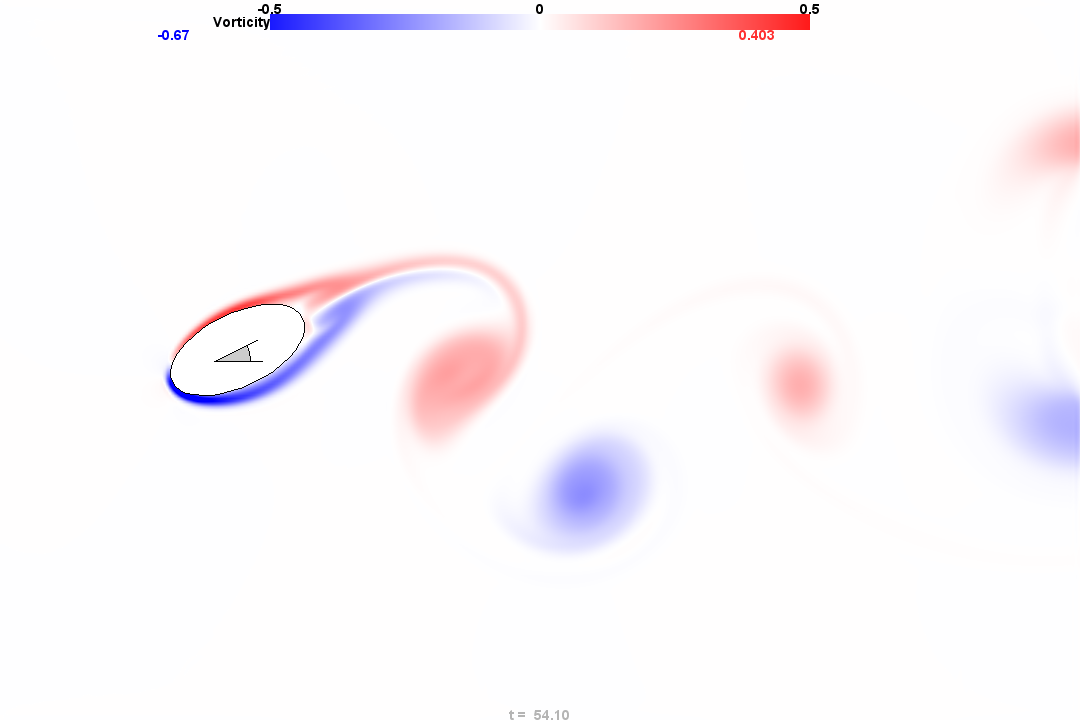}}
	\end{minipage}
	\begin{minipage}{0.4\textwidth}
		\scalebox{1}[-1]{\includegraphics[trim=4cm 4cm 2cm 4cm, clip=true, 
		width=\textwidth]{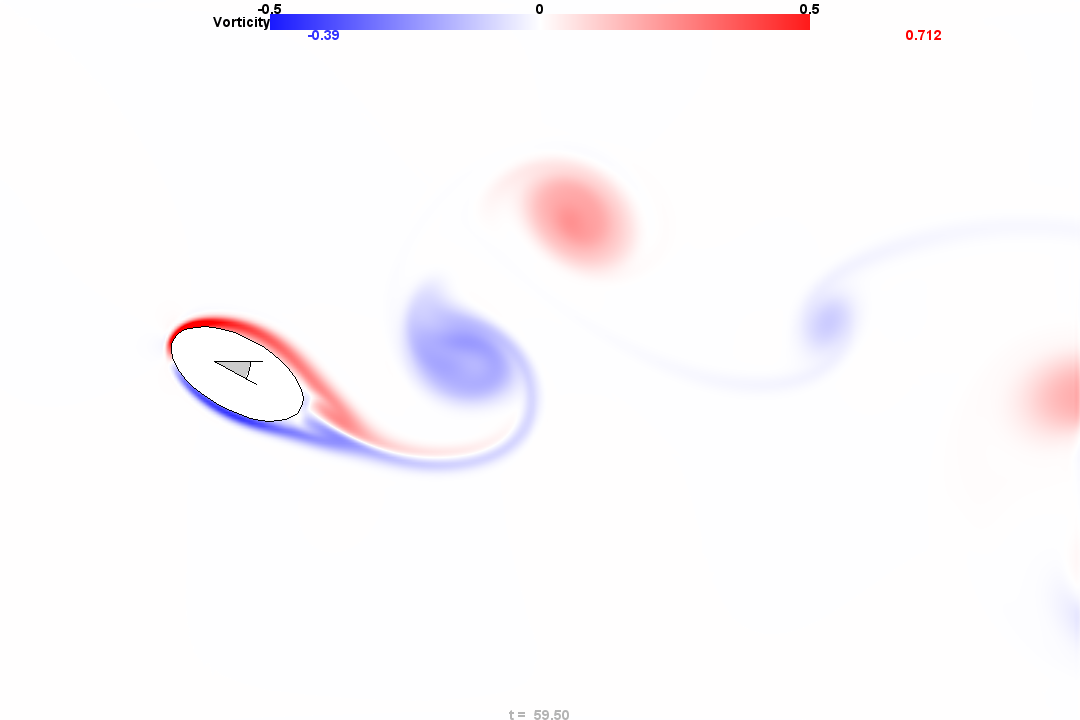}}
	\end{minipage}
	\rotatebox[origin=c]{270}{
	\begin{minipage}{26mm}\centering
	\begin{snugshade} 
		\scalebox{0.8}{$r/L = 0.18$ }
	\end{snugshade}
	\end{minipage}}\\
	\begin{minipage}{0.4\textwidth}
		\scalebox{1}[-1]{\includegraphics[trim=4cm 3cm 2cm 3cm, clip=true, 
		width=\textwidth]{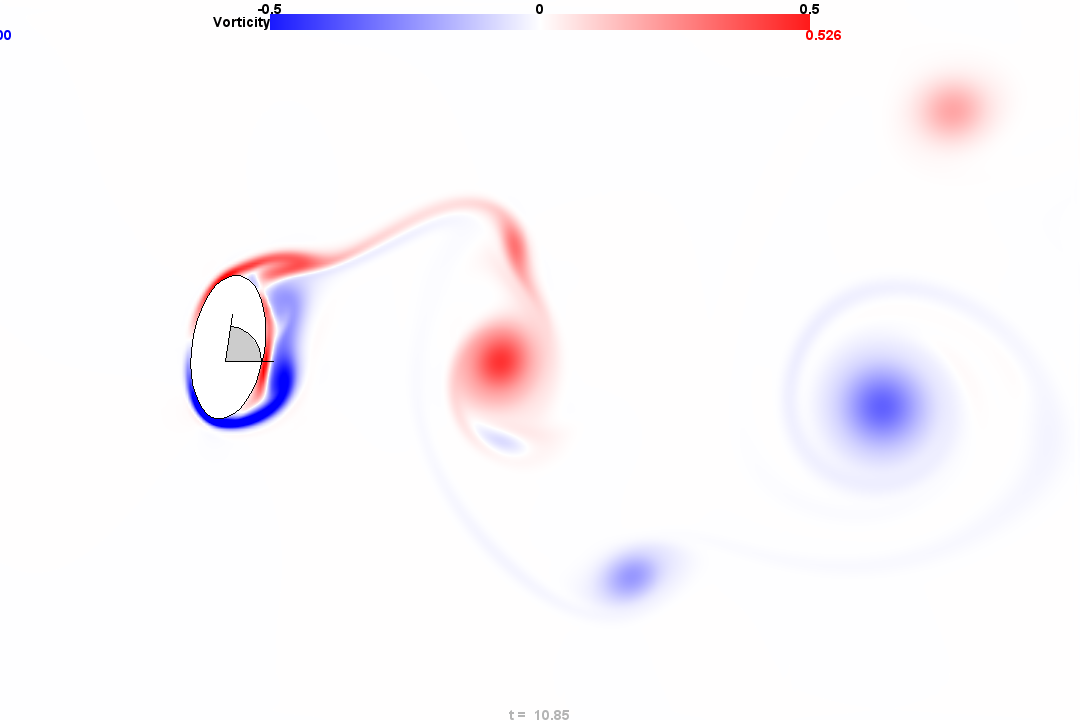}}
	\end{minipage}
	\begin{minipage}{0.4\textwidth}
		\scalebox{1}[-1]{\includegraphics[trim=4cm 3cm 2cm 3cm, clip=true, 
		width=\textwidth]{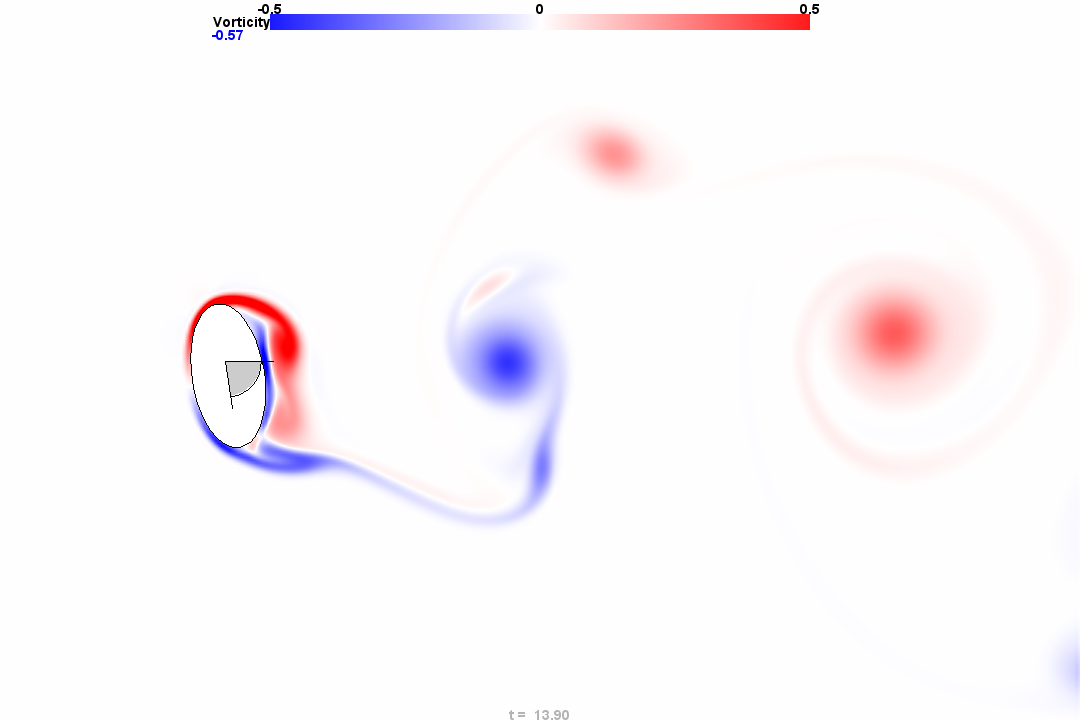}}
	\end{minipage}
	\rotatebox[origin=c]{270}{
	\begin{minipage}{29mm}\centering
	\begin{snugshade} 
		\scalebox{0.8}{$r/L = 0.1$ }
	\end{snugshade}
	\end{minipage}}\\
	\begin{minipage}{0.4\textwidth}
		\scalebox{1}[-1]{\includegraphics[trim=4cm 2cm 2cm 6cm, clip=true, 
		width=\textwidth]{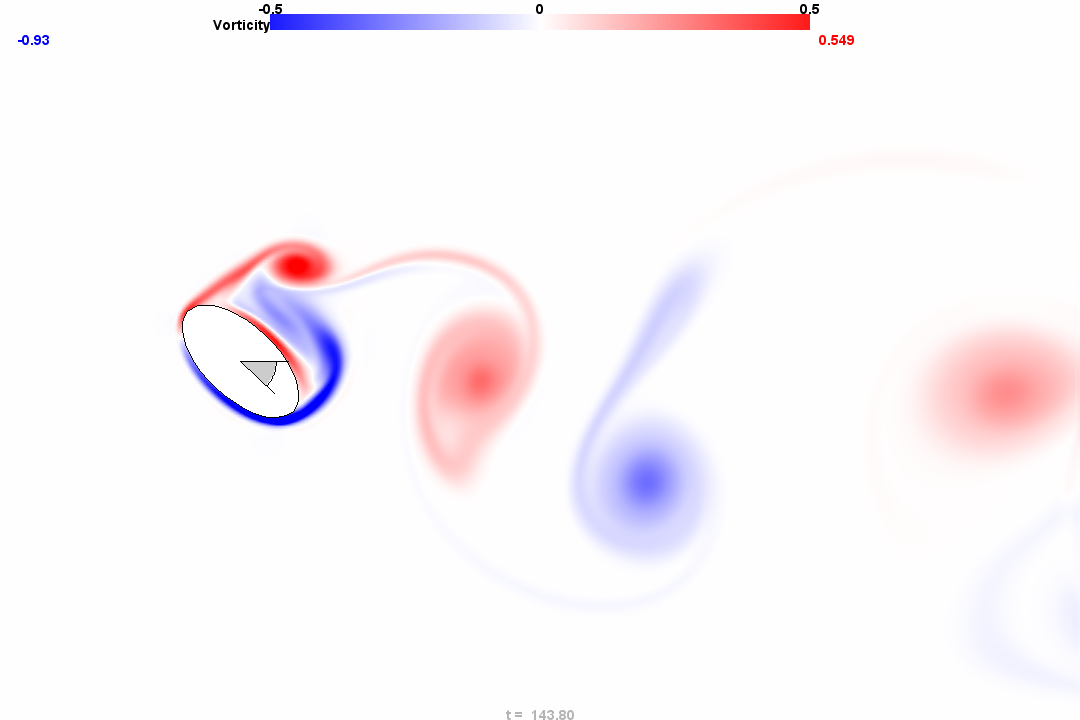}}
	\end{minipage}
	\begin{minipage}{0.4\textwidth}
		\scalebox{1}[-1]{\includegraphics[trim=4cm 2cm 2cm 6cm, clip=true, 
		width=\textwidth]{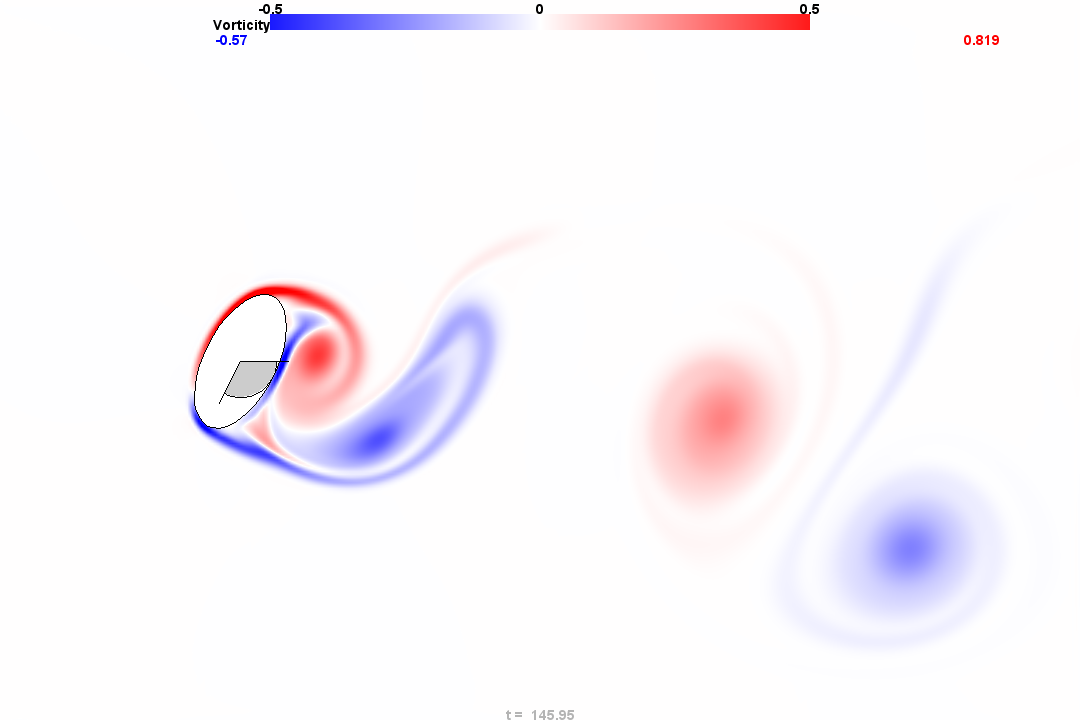}}
	\end{minipage}
	\rotatebox[origin=c]{270}{
	\begin{minipage}{26mm}\centering	
	\begin{snugshade} 
		\scalebox{0.8}{$r/L = 0$ }
	\end{snugshade}
	\end{minipage}}\\
	\caption{Vorticity field (gray, red/blue online) in the wake of the ellipse (black outline, travelling right-to-left) as it rotates freely in limit cycle motion at $Re=1000$. Two times are shown for each value of $r$ at approximately the extreme values of $\phi$. The instantaneous value of $\phi$ is indicated on the pivot of the ellipse for reference.}
	\label{fig: periodic wake}
\end{figure}

\begin{figure}
	\centering
	\includegraphics[]{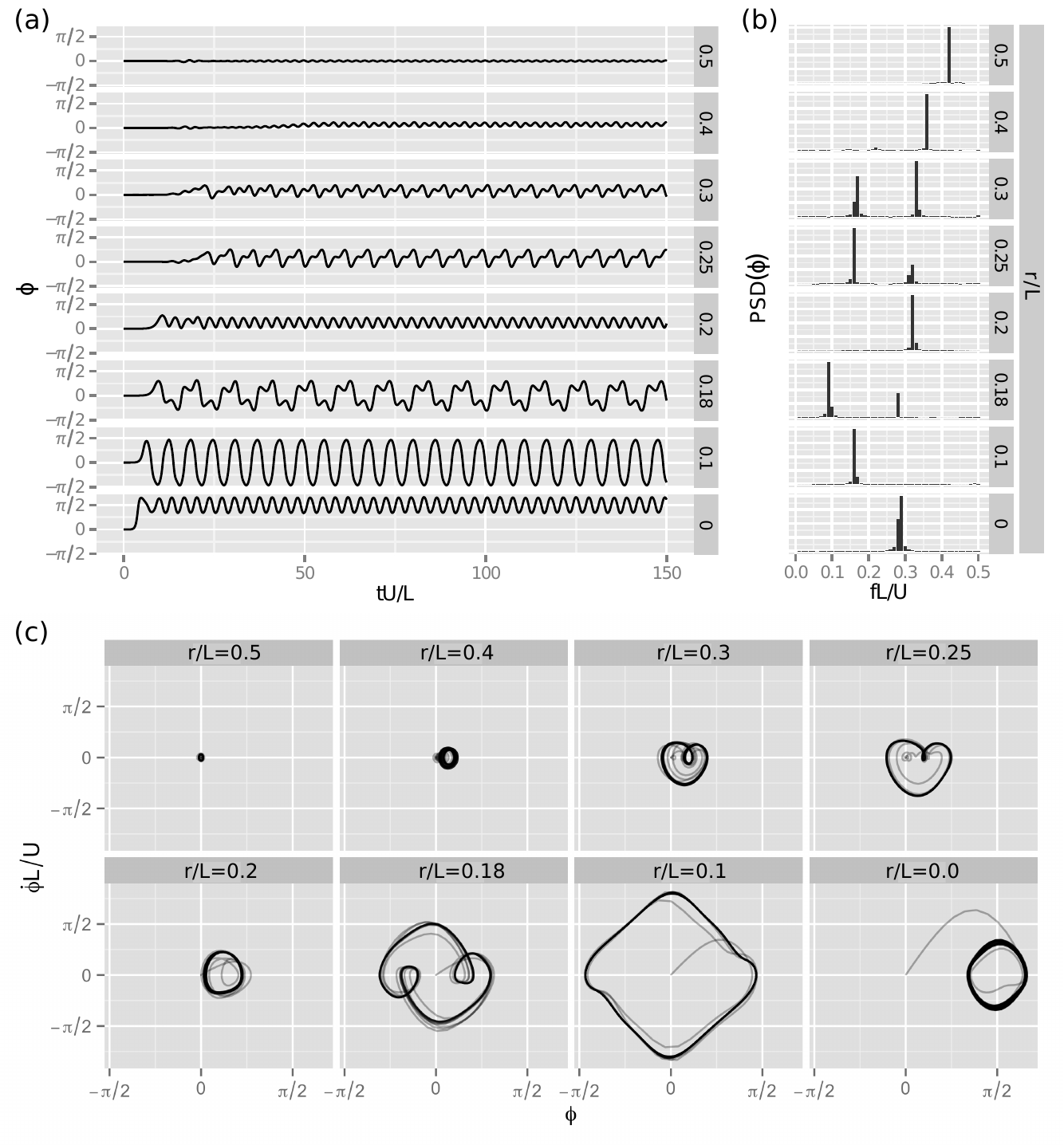}
	\caption{(a) Time history, (b) normalized power spectral density, and (c) phase portraits of $\phi$ for values of $r/L$ which produce periodic limit cycles at $Re=1000$. The phase portraits are drawn with 70\% transparency so that overlapping orbits appear darker.}
	\label{fig: periodic history}
\end{figure} 

The limit cycle behaviour is detailed first. Figure \ref{fig: periodic wake} shows the vortex wake and Fig \ref{fig: periodic history} the time history, power spectral density, and phase portraits of the rotation angle $\phi$ for values of $r/L$ which result in limit cycle trajectories. A fair understanding of the basic dynamics of the system is achieved by combining the structural state information in Fig \ref{fig: periodic history} with the fluid state information in Fig \ref{fig: periodic wake}. 

When the lever arm is large, the torque induced by the bluff-body drag limits the motions to small amplitude and speed. At $r/L=0.5$ the limit cycle is centered on $\phi=0$ but as $r/L$ is decreased the Hopf bifurcation in Eq~\ref{eq: drag torque} is crossed some time before $r/L=0.4$, destabilizing the $\phi=0$ limit cycle and creating two stable limit cycles on either side. This indicates that the value of $\beta$ is approximately $1/2$ for this Reynolds number. The figures have been oriented such that the trajectories all move to the positive limit cycle, but adjusted initial conditions make either branch accessible.

There is a near-periodic orbit at $r/L=0.4$ with a wake characterized by a standard Karman vortex street. In the literature of vortex induced vibrations \citep{Williamson2004} this is categorized as a 2S wake, having two single vortices per cycle. Before $r/L=0.3$, the trajectories undergo period doubling as seen clearly in the frequency content of $\phi$. This wake features two pairs of vortices, a 2P wake, with one pair stronger than the other. A second period-2 limit-cycle appears at $r/L=0.25$ which has 20\% larger amplitude and a dominant period-2 component. The wake for this cycle is similar, but one of the vortices in the smaller pair is grouped with the larger pair leaving the other small vortex isolated, making this a T+S wake (Triplet plus Single). 

The region from $0.25>r/L>0.15$ features mostly non-periodic trajectories, but a small amplitude period-1 limit cycle centered on $\pi/8$ is found at $r/L=0.2$ with an asymmetric P wake. A periodic response is also found at $r/L=0.18$ with a large period-3 limit cycle which alternates between the positive and negative branches from the fixed-point analysis in \S\ref{formulation}. This is the first wake which features leading edge vortex (LEV) separation and these vortices power the trajectory between the two branches, joining the positive and negative orbits. The wake is a symmetric version of the $r/L=0.25$ limit cycle, with a 2P+2S wake. 

A stable limit cycle is found between $0.15>r/L>0.05$ with large amplitude ($\Phi\simeq\pi/2$) and high velocity ($\dot\Phi\simeq \frac 34 \pi U/L$) as in rotational galloping. The frequency content shows that this a period-2 limit cycle with nearly zero period-1 component. LEVs form a vortex street on the centerline with peak vorticity levels double that of the $r/L=0.4$ limit cycle wake. Smaller trailing edge vortices form off-center between the LEVs resulting in a 4S wake. As $r/L\rightarrow0$  this large amplitude orbit decays intermittently into a smaller period-1 cycle centered on $\phi=\pm\pi/2$ with an asymmetric P wake.

Overall, the vortex shedding frequency is completely dominant for the limit cycle cases. The number of vortices corresponds directly to the period of the cycle; two for the period-1 cycles (2S, P), four for the period-2 cycles (2P,T+S), and six for the period-3 cycle (2P+2S). The period-1 frequency of the trajectories is also determined by the vortex dynamics. At $r/L=0.5$, $\phi_s^*=0$, and the cross-stream width is $w=L/2$, which results in the standard bluff body Strouhal number $St=fw/U\simeq 0.2$. As $r/L$ is decreased, the trajectories spread from $\phi=0$ increasing the effective cross stream width, accounting for the decreased period-1 frequency observed in Fig~\ref{fig: periodic history}.

\begin{figure}
	\centering
	\includegraphics[width=\textwidth]{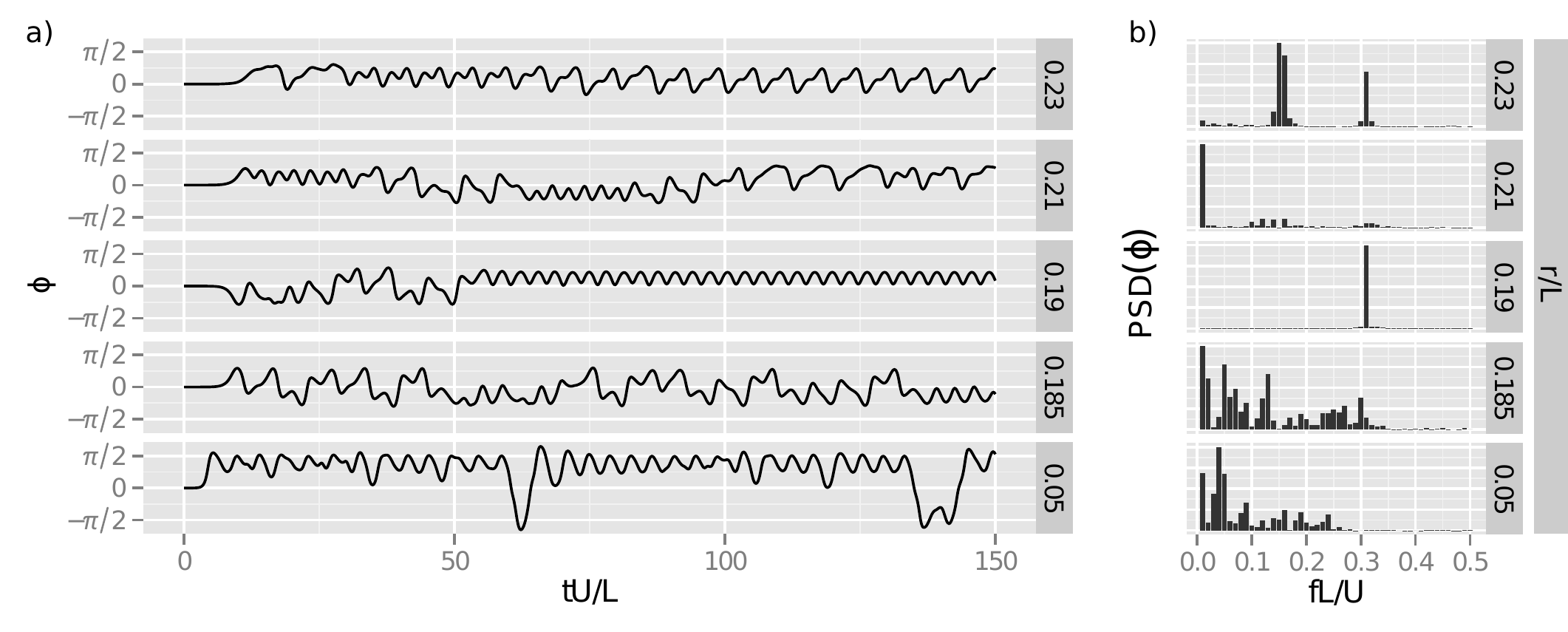}
	\caption{(a) Time history and (b) normalized power spectral density of $\phi$ for values of $r/L$ which produce orbits with intermittently periodic behaviour at $Re=1000$.}
	\label{fig: int history}
\end{figure}

Non-periodic behaviour is found between most of these periodic limit cycles for $r/L<0.3$. Figure \ref{fig: int history} shows the response for selected  trajectories which display many types of non-periodic behaviour associated with bistable oscillators. The trajectory at $r/L=0.23$ transition from one limit cycle to another (the $r/L=0.3$ cycle to the $r/L=0.25$ cycle). $r/L=0.19$ shows an extended non-periodic phase followed by stable period-1 motion. The branch the trajectory eventually oscillates around is sensitive to the initial conditions. The other three ($r/L=0.21$, 0.185, and 0.05) show windows of periodic behaviour interspersed with non-periodic bursts of activity. Many of these trajectories feature non-periodic transitions from one branch to the other which results in a broadband frequency response.

\begin{figure}
	\centering
	\subfloat[]{
		\includegraphics[width=0.67\textwidth]{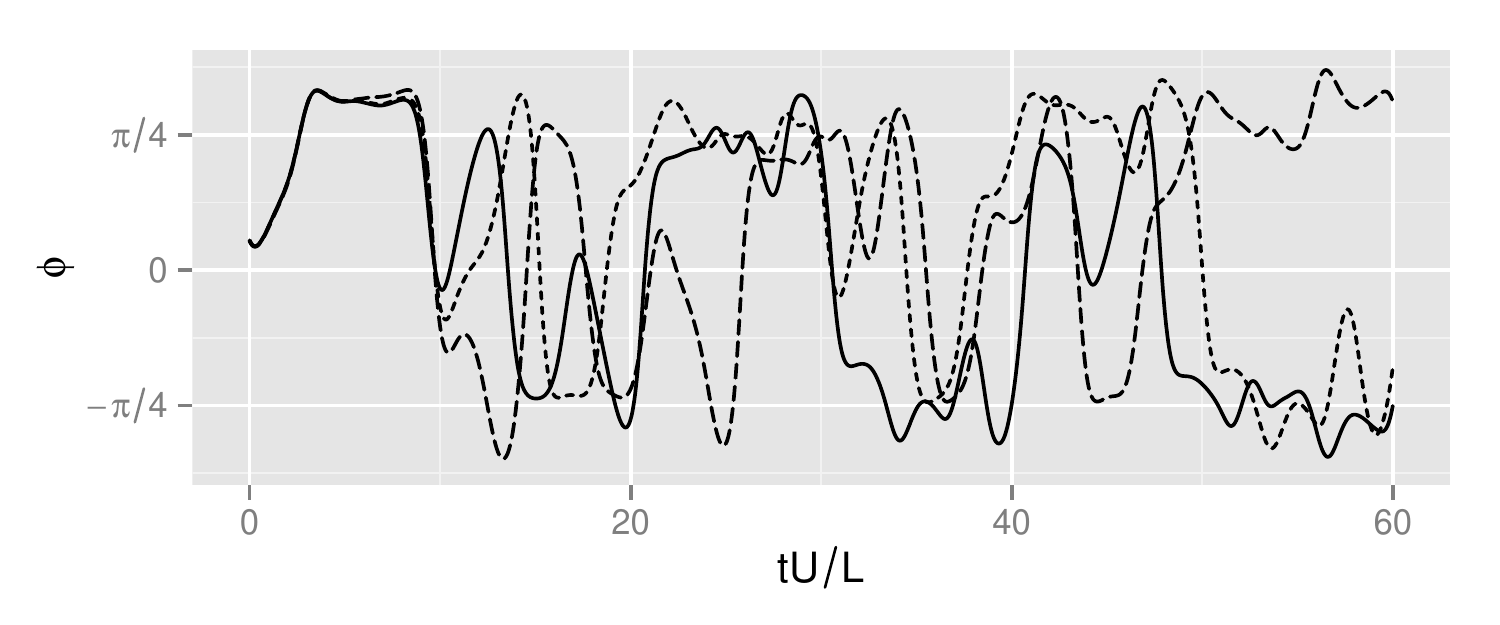}} 
	\subfloat[]{
		\includegraphics[width=0.33\textwidth]{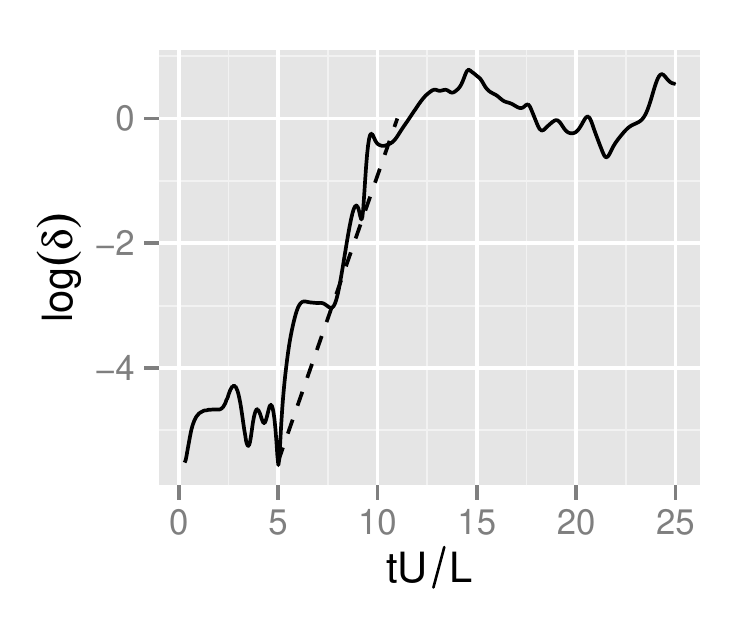}
		\label{fig: chaos del}} \\
	\subfloat[]{
		\includegraphics[width=0.4\textwidth]{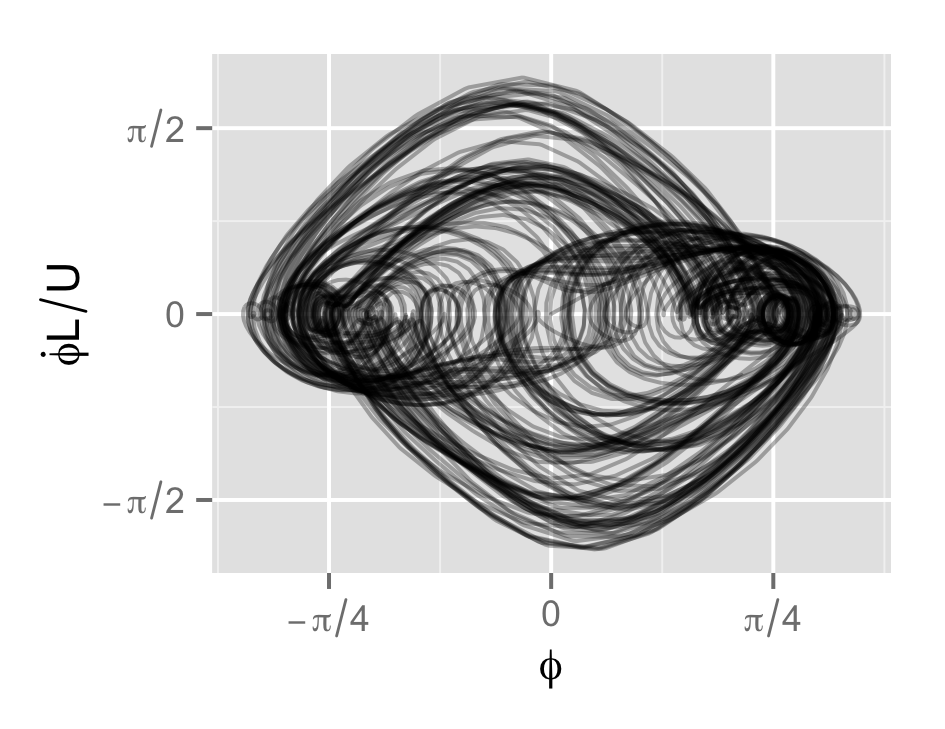}
		\label{fig: chaos phase}} 
	\subfloat[]{
		\includegraphics[width=0.3\textwidth]{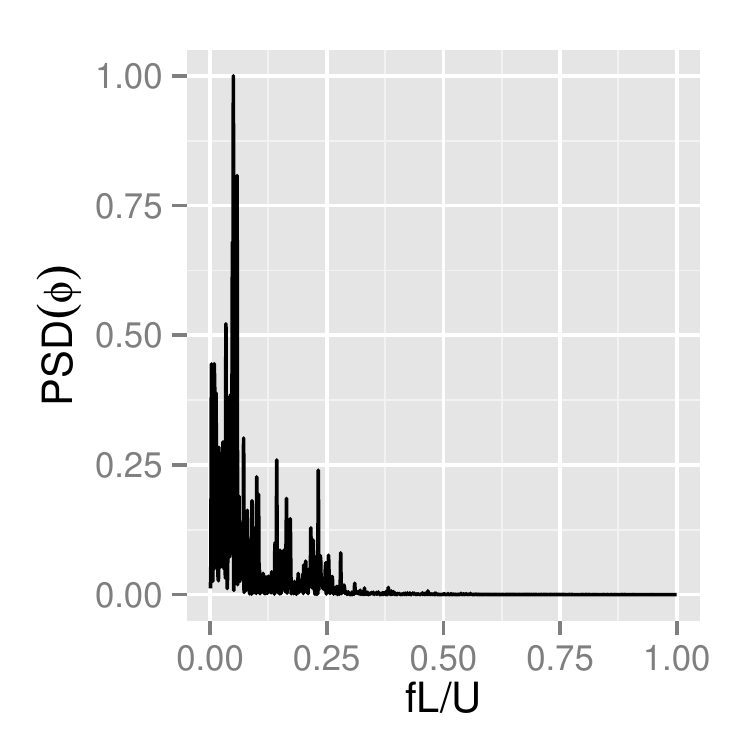}
		\label{fig: chaos freq}} 
	\subfloat[]{
		\includegraphics[width=0.3\textwidth]{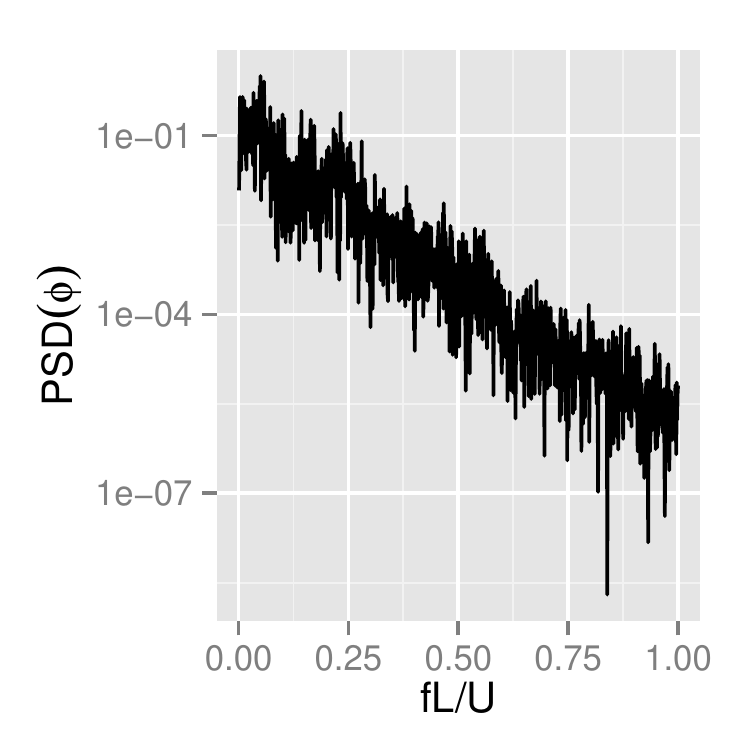}}
	\caption{Trajectory characteristics of chaotic rotations with $r/L=0.16$ and $Re=1000$. (a) Time history of three trajectories started with initial conditions $\delta_0=.25\%$ apart. (b) The evolution of the average distance between the trajectories in (a). The slope of the dashed line is 0.9 for reference. (c) Phase portrait for one trajectory from $0<tU/L<10^3$ using 70\% line transparency. (d) Linear and (e) log plot of the normalized power spectral density of $\phi$ for the same trajectory as in (c). }
	\label{fig: chaos}
\end{figure}

A strong fully chaotic response is found in the window between the large period-3 orbit at $r/L=0.18$ and period-2 orbit at $r/L=0.1$. The chaotic response for $r/L=0.16$ is examined in detail in Fig \ref{fig: chaos}. The time history is shown for a set of trajectories with a less than 1\% distance between their initial conditions. Figure \ref{fig: chaos del} shows the average evolution of the distance between trajectories. The solutions remain close for times $tU/L<5$ while the flow around the cylinder initially develops. As the starting vortices detach from the ellipse the distance between the trajectories grows exponentially, with a (largest) Lyapunov exponent of 0.9. By $tU/L=12$ the trajectories have completely diverged. Figure \ref{fig: chaos phase} shows the phase portrait for one trajectory for $0\leq tU/L\leq 10^3$. Similarities to the limit-cycles found at $r/L=0.1,0.18$ are evident, but most of the structural states within the envelope are visited at least once. The most common states are at ($\phi=\pm\pi/4$, $\dot\phi=0$) which are the approximate fixed points from Fig~\ref{fig: spring} given $\beta=0.5$. Fig~\ref{fig: chaos} also plots the normalized power spectral density of the trajectories. There is a peak at $fL/U=0.05$, which corresponds to period-6, but the response is generally broadband. 

\begin{figure}
	\centering
	\subfloat{
		\scalebox{1}[-1]{\includegraphics[trim=4cm 6cm 2cm 4cm, clip=true, 
		width=0.4\textwidth]{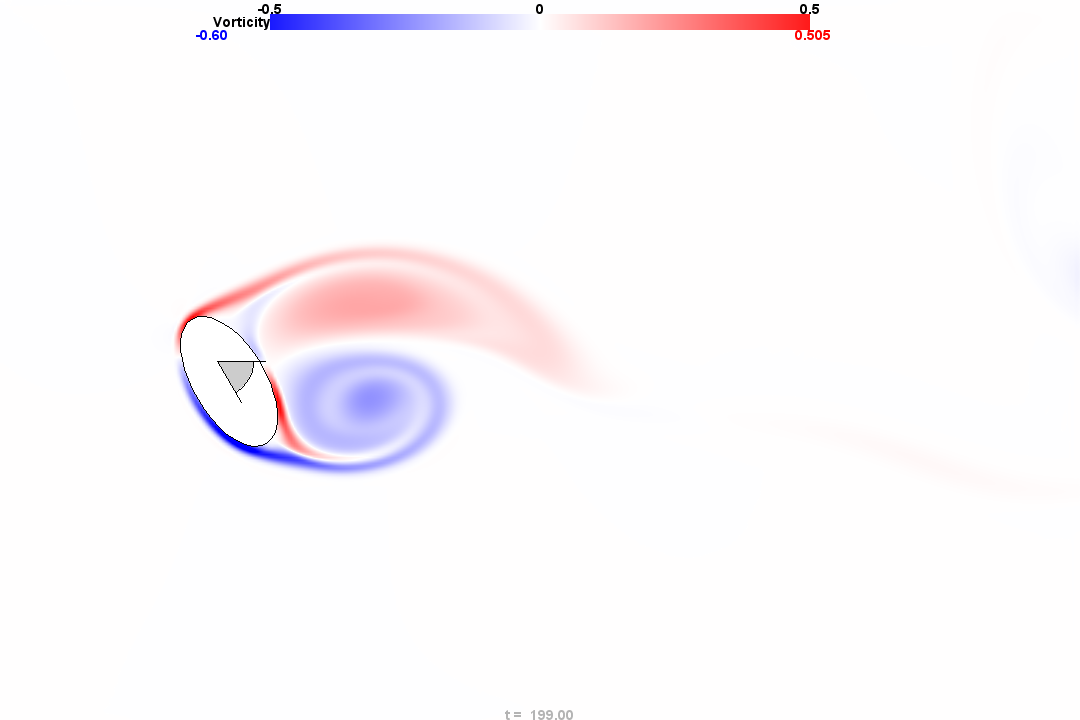}}}
	\subfloat{
		\scalebox{1}[-1]{\includegraphics[trim=4cm 6cm 2cm 4cm, clip=true, 
		width=0.4\textwidth]{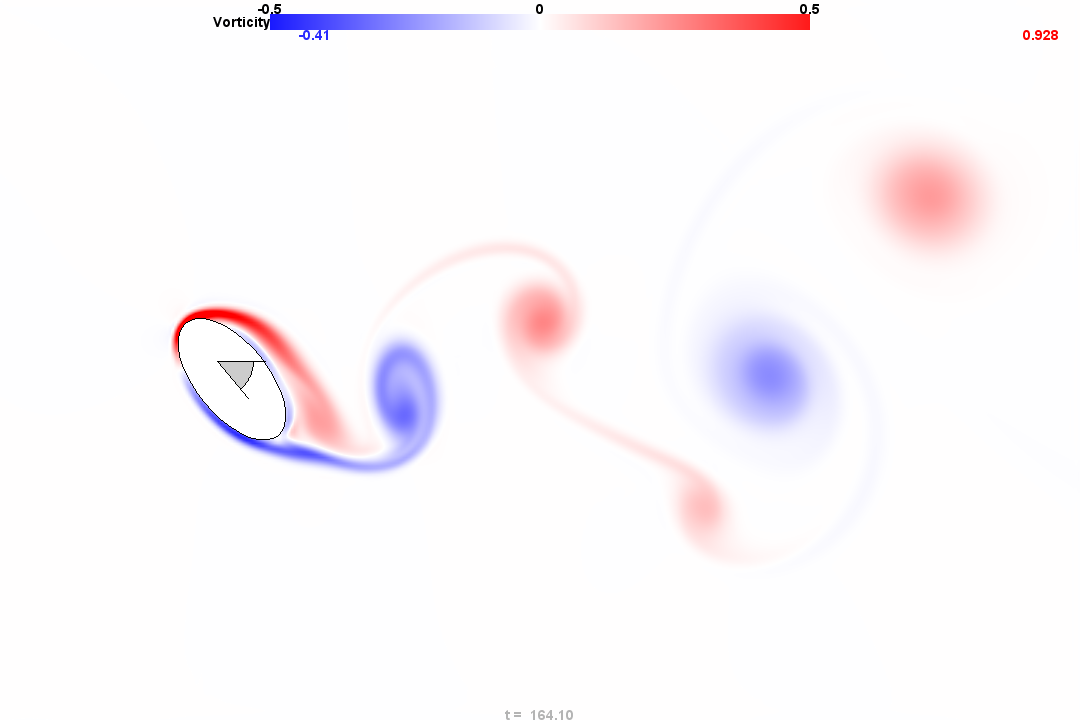}}} \\
	\subfloat{
		\scalebox{1}[-1]{\includegraphics[trim=4cm 4cm 2cm 4cm, clip=true, 
		width=0.4\textwidth]{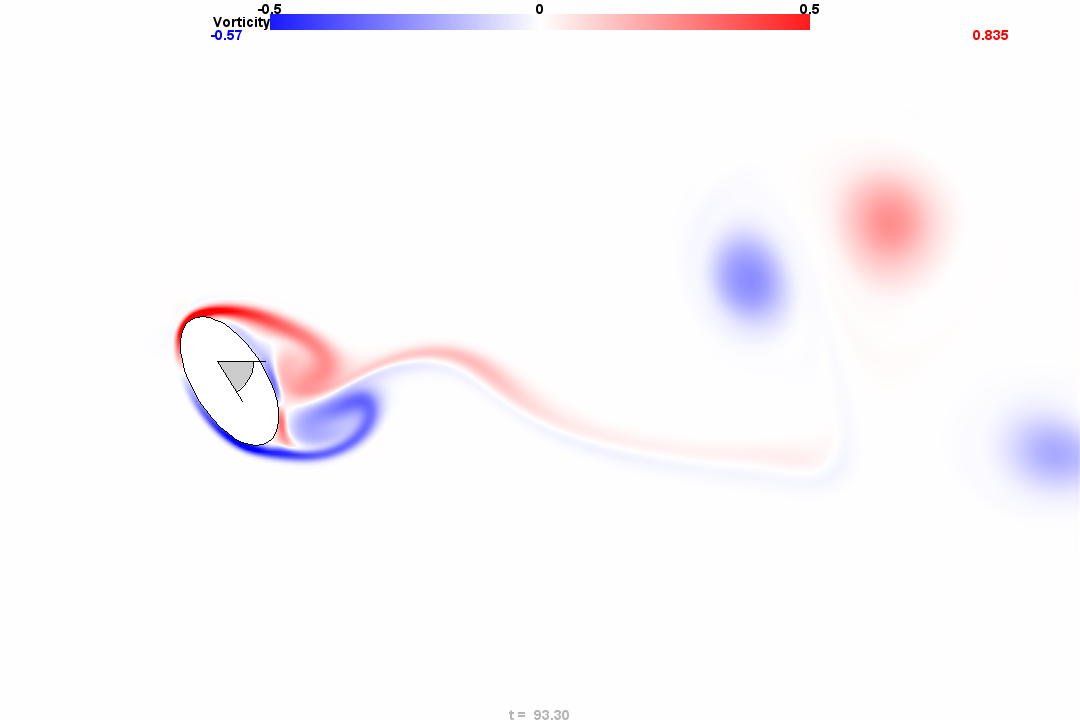}}}
	\subfloat{
		\scalebox{1}[-1]{\includegraphics[trim=4cm 4cm 2cm 4cm, clip=true, 
		width=0.4\textwidth]{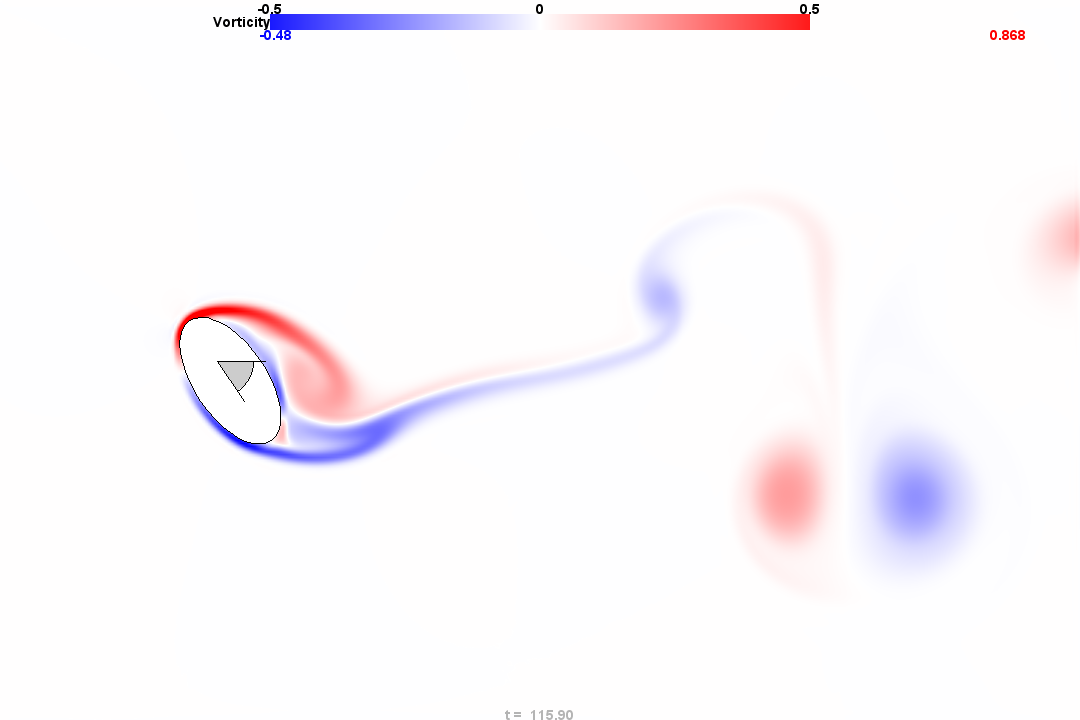}}}
	\caption{Vorticity field (gray, red/blue online) in the wake of the ellipse (black outline, travelling right-to-left) as it rotates freely in chaotic motion with $r/L=0.16$ and $Re=1000$. Four times are shown at approximately the same values of $\phi$ and $\dot\phi$.}
	\label{fig: chaos wake}
\end{figure}

Figure \ref{fig: chaos wake} shows the vortex wake for the same chaotic case. In this figure, four times are shown when the trajectories pass through  approximately the same point in the phase space. However, the wake in each image is completely unique; a lone pair of bound vortices,  a set of 2P+S vortices, and an isolated vortex pair shooting far off centerline upwards or downwards. The torque applied to the ellipse is unique in each case, leading to completely different future trajectories.

\begin{figure}
	\centering
	\subfloat[ ]{
		\includegraphics[width=0.33\textwidth]{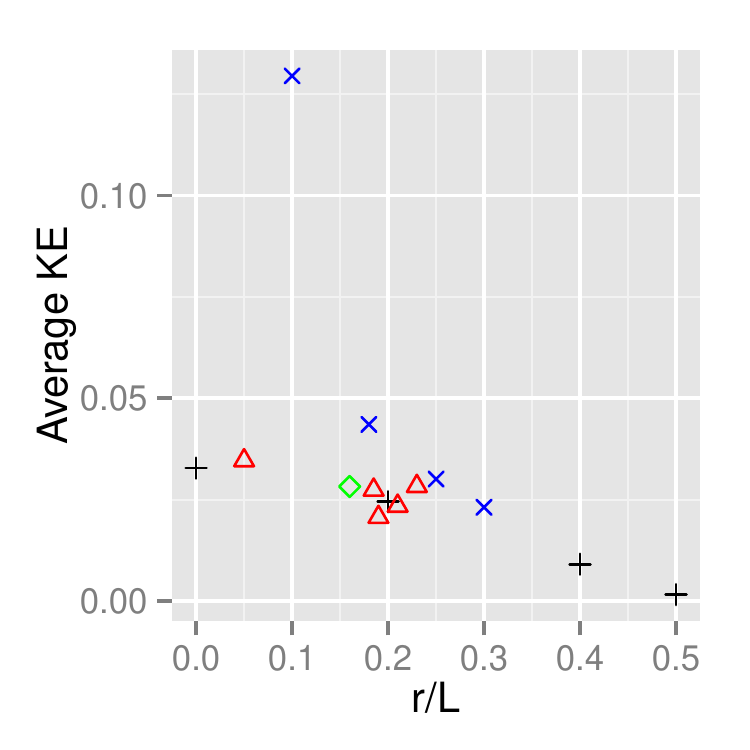}}
	\subfloat[ ]{
		\includegraphics[width=0.33\textwidth]{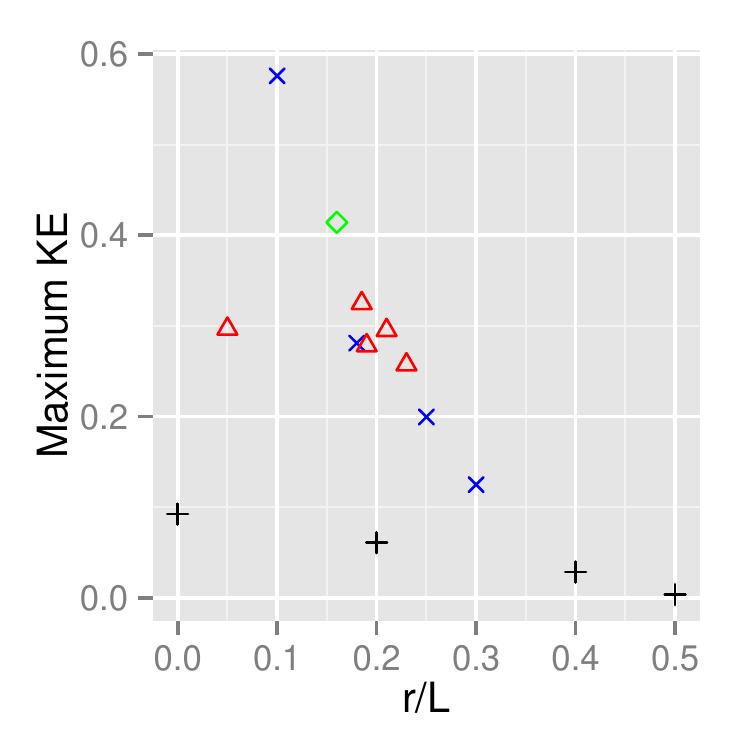}}
	\subfloat[ ]{
		\includegraphics[width=0.33\textwidth]{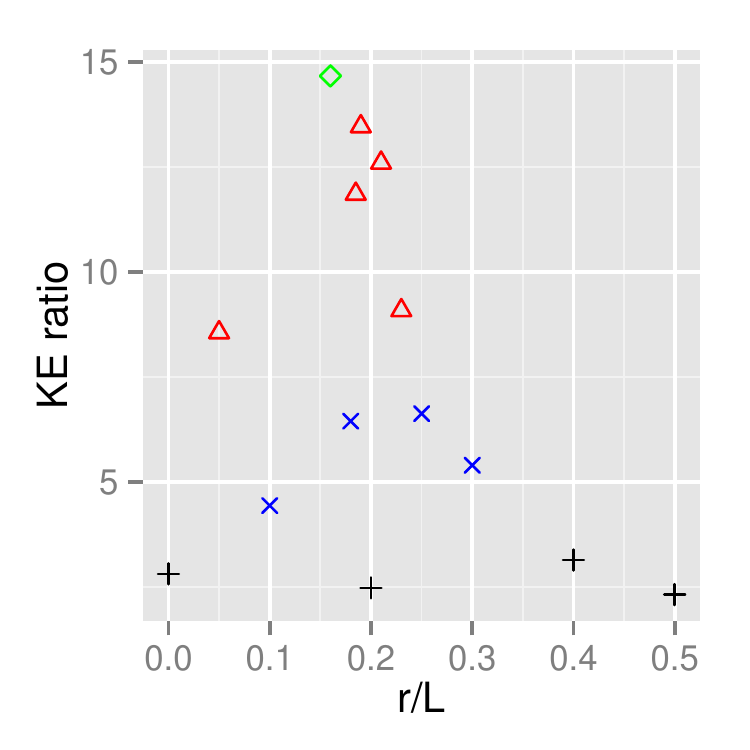}}
	\caption{(a) Average and (b) Maximum rotational kinetic energy $\frac 12 I_0 \dot\phi^2$ of the ellipse scaled by the constant translational kinetic energy. (c) Ratio of the maximum to the average energy. Black crosses: period-1 cases, Blue x's: higher period limit cycles, Red triangles: intermittent cases, Green diamond: chaotic $r=0.16L$ case. $Re=1000$.}
	\label{fig: ke}
\end{figure}

The rotational kinetic energy of the ellipse during free oscillations is shown in Fig \ref{fig: ke} as a function of $r/L$. The trend is an increase in KE with decreasing $r/L$ until the $r/L=0.1$ cycle is reached, which has the maximum value observed. However, the details are markedly different for the different types of trajectories. The limit cycle cases show average energy levels that follow the trend mentioned, but the period-1 limit cycles have very low maximum energy levels. In contrast the average kinetic energies of the non-periodic cases are essentially constant, but their maximum energy level increases with the trend. Plotting the ratio of maximum to average rotational kinetic energy cleanly divides the period-1 cycles below 3, the non-periodic orbits above 7, and the higher period orbits between. The $r/L=0.16$ chaotic case has the maximum ratio of nearly 15.

\begin{figure}
	\centering
	\includegraphics[trim=0 1.5cm 0 0, clip=true, width=\textwidth]{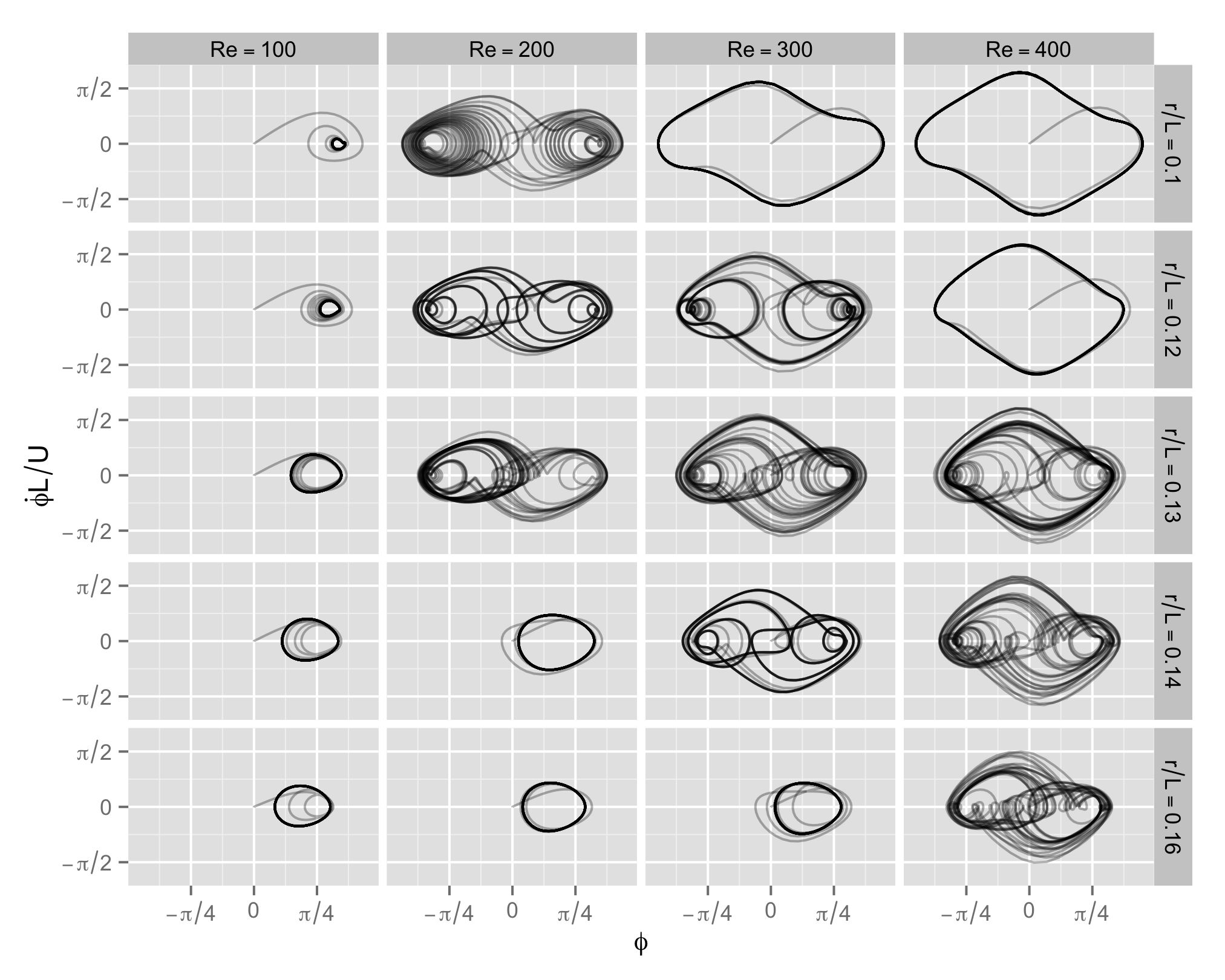}
	\caption{Phase portraits for $Re=100-400$ and $r/L=0.1-0.16$. Figure \ref{fig: chaos phase} shows the trajectory for $r/L=0.16$ and $Re=1000$.}
	\label{fig:re}
\end{figure}

\begin{figure}
	\centering
	\includegraphics[trim=0 1.5cm 0 0, clip=true, width=\textwidth]{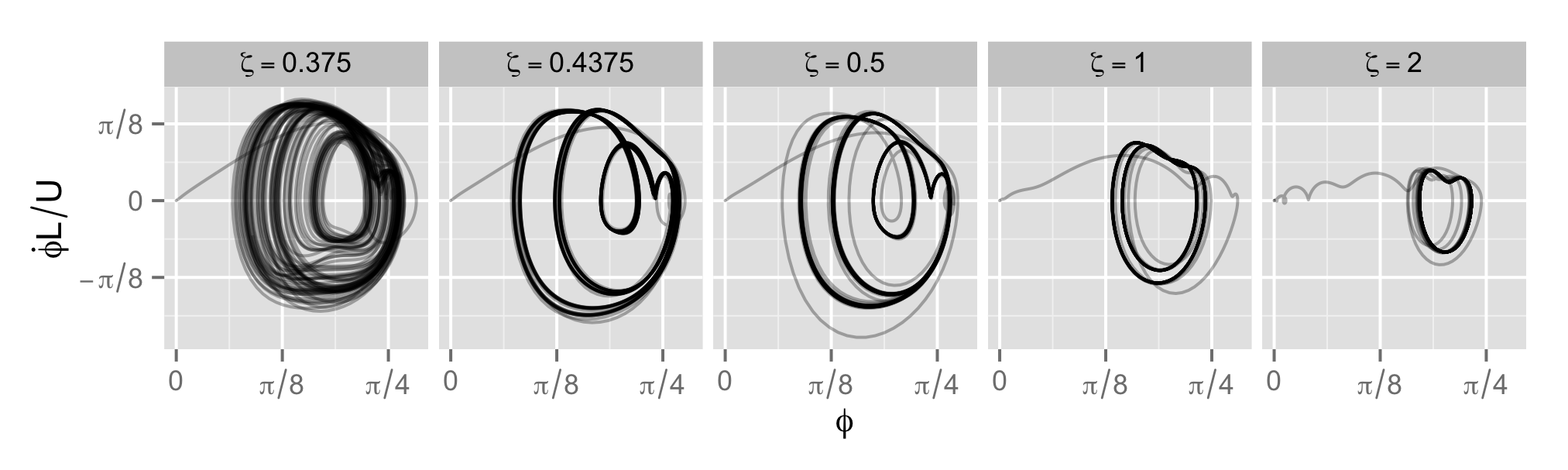}
	\caption{Phase portraits for $\zeta=0.375-2$ with $Re=1000$ and $r/L=0.16$. Figure \ref{fig: chaos phase} shows the trajectory for the undamped case.}
	\label{fig:zeta}
\end{figure}

Finally, the dependence on damping is investigated in two ways; first, through the reduction of Reynolds number, and second, through the introduction of linear structural damping. Figure~\ref{fig:re} shows the behaviour of the system as the Reynolds number is reduced for the highest energy trajectories, $0.1\leq r/L \leq 0.16$. The increased fluid damping reduces the size and speed of oscillations throughout this range, and at $Re=100$ all trajectories have been damped to period-1 limit cycles around a single branch from the fixed point analysis in \S\ref{formulation}. However, there are still strong non-periodic responses found for Reynolds numbers as low as $200$. The change in Reynolds number also adjusts the location of the peak chaotic response from $r/L=0.16$ (as in Fig~\ref{fig: chaos phase}) down to $r/L\sim0.1$. This is likely due to a change in the balance of the magnitude of the Munk moment and drag induced moment, quantified by the parameter $\beta$ in Eq~\ref{eq: drag torque}. The increased fluid damping has also stabilized higher period responses than were observed at $Re=1000$, with the most extreme example being a period-14 limit cycle found at $Re=200,\ r/L=0.12$.

Figure \ref{fig:zeta} shows the result of adding linear structural damping torque to the system for the case of $r/L=0.16$ and $Re=1000$ which is chaotic for the undamped system. The damping coefficient $\zeta$ is nondimensionalized by $I_0 f_1/\pi$ where $f_1L/U=0.3$ is the observed period-1 frequency in this range. Therefore, $\zeta=1$ corresponds to critical damping of this frequency in the absence of fluid excitation. As with the Reynolds number study, adding damping results in a less energetic trajectory, and $\zeta=2$ (overdamped) results in a period-1 limit cycle around a single branch. Less extreme damping destabilizes this trajectory and leads to period-2 cycle at $\zeta=1$, period-3 at $\zeta=0.5$, and a period-6 cycle at $\zeta=0.4375$. By $\zeta=0.375$ the trajectory is fully chaotic but much less energetic than the undamped case, with velocities 1/4 the magnitude and no switching between the positive and negative branches.

\section{Discussion and Conclusions}\label{conclusion}

This work considers a simple towed elliptical cylinder which is free to rotate around a pivot, and demonstrates that the system responds with a rich set of dynamics, including limit cycles, period doubling, and chaotic motions. The mean restoring torque was approximated analytically and used to show that the system has a region of natural bistability dependant on the distance $r/L$ between the centroid and the pivot. The underlying system is closely related to forced bistable oscillators, which are well known to incorporate these dynamical behaviours. 

The ellipse motion in the limit cycles cases is dominated by the vortex dynamics, as observed in the literature of autorotation \citep{Lugt1983}. The frequency and number of oscillations in each limit cycle corresponds directly to the bluff body Strouhal number, indicating that this is the dominant time-scale in the system (Fig~\ref{fig: periodic history}). Decreasing $r/L$ from 0.2 to 0.1 changes the character of the wake from trailing edge to leading edge vortex shedding with a corresponding increase in the amplitude of motion and the occurrence of trajectories visiting both of the stable branches (Fig~\ref{fig: periodic wake}). 

The vortex wake of the chaotic case at $r/L=0.16$ is only weakly correlated with the state of the ellipse, featuring any number of vortices as well as pairs of vortices shooting far off center-line in either direction (Fig~\ref{fig: chaos wake}). This sensitive wake history effect is a necessary condition for chaotic motions and stands in sharp contrast to the period-1 oscillations before the Hopf bifurcation in which the wake is fully determined by the instantaneous state of the ellipse. The irregularity of the wake also illustrates the limits of the analogy to the classic forced bistable systems such as \cite{Holmes1979}. The current system is self-exciting, not subject to a harmonic forcing. Therefore there is no fixed frequency or amplitude of forcing, and a constant period Poincare map is no better at clarifying the strange attractor than a random sample of the phase portrait.

The $r/L=0.1$ period-2 limit cycle has the largest rotational energy observed in this study, with a maximum of nearly 60\% of the translational kinetic energy. The ratio of maximum to average kinetic energy perfectly separates the trajectory types, consistent with the cascade of energy into a wider frequency spectrum. The ratio for the chaotic case is nearly 15, indicating extremely nonlinear loading on the ellipse in this trajectory. This is consistent with the `snap' loading observed in chaotic FSI problems with flexible bodies \citep{Sipcic1990,Connell2007}.

The energy of the ellipse motion is reduced as the damping losses are increased either through structural damping or decreased Reynolds number. In this way it is possible to restrict even the highest energy cases ($0.1\leq r/L \leq 0.16$) to low-amplitude period-1 limit cycles. Fluid damping is seen to stabilize the motion of the system, and at $Re=200$ a period-14 limit cycle is observed, as well as fully non-periodic motion. Similarly, moderate structural damping enables period-6 limit cycles, before finally degenerating to chaotic motion for $\zeta\leq 0.375$.

The damping studies demonstrate that the energy of the FSI system is a critical factor for the stability of low-period limit cycles. This can be extended to the undamped case by considering how the energy changes as $r/L$ is adjusted. As $r/L$ is reduced from 0.5 to 0, the trajectories diverge from $\phi=0$ increasing the effective cross-stream width of the object. This has the combined effect of decreasing the bluff body shedding frequency and, crucially, of increasing the mean drag force on the tow point. The tow point is therefore doing more work on the FSI system, and the total KE must increase with the decrease in $r/L$. The system cannot accommodate this energy by increasing its rotation frequency because this is set from the Strouhal number. Nor can the trajectories visit $\phi$ much greater than the stable fixed points established in \S\ref{formulation} because of the extremely large restoring force beyond this point (Fig~\ref{fig: torque}). The only resolution is for the system to release this increased energy into infrequent high-energy vortex shedding events. This leads to the observed period doubling, increased complexity of the wake, `snap' loading, and the onset of chaos.

New applications such as sensors and energy extraction devices rely on the sensitive response of the structure to the flow. This work successfully demonstrates that even the simplest possible nonlinear FSI problem incorporates the full range of sensitive chaotic responses, and the range of possible dynamics only increases as variations to the structural stiffness, mass ratio, and body cross-section are considered. As such, it may serve as a canonical example of nonlinear interactions, suitable for education, detailed analytic and experimental analysis, and to instigate new passive or minimally-actuated hydrodynamic sensor and energy extraction designs.

\bibliographystyle{jfm}
\bibliography{../../complete}

\begin{thebibliography}{28}
\expandafter\ifx\csname natexlab\endcsname\relax\def\natexlab#1{#1}\fi

\bibitem[Abdelkefi {\em et~al.\/}(2013)Abdelkefi, Hajj \&
  Nayfeh]{Abdelkefi2013}
{\sc Abdelkefi, A, Hajj, MR \& Nayfeh, AH} 2013 Piezoelectric energy harvesting
  from transverse galloping of bluff bodies. {\em Smart Materials and
  Structures\/} {\bf 22}~(1), 015014.

\bibitem[Alonso {\em et~al.\/}(2010)Alonso, Meseguer, Sanz-Andres \&
  Valero]{Alonso2010}
{\sc Alonso, G., Meseguer, J., Sanz-Andres, A. \& Valero, E.} 2010 On the
  galloping instability of two-dimensional bodies having elliptical
  cross-sections. {\em Journal of Wind Engineering and Industrial
  Aerodynamics\/} {\bf 98}~(8-9), 438 -- 448.

\bibitem[Arrieta {\em et~al.\/}(2010)Arrieta, Hagedorn, Erturk \&
  Inman]{Arrieta2010}
{\sc Arrieta, AF, Hagedorn, P, Erturk, A \& Inman, DJ} 2010 A piezoelectric
  bistable plate for nonlinear broadband energy harvesting. {\em Applied
  Physics Letters\/} {\bf 97}~(10), 104102--104102.

\bibitem[Barrero-Gil {\em et~al.\/}(2010)Barrero-Gil, Alonso \&
  Sanz-Andres]{Barrero2010}
{\sc Barrero-Gil, A., Alonso, G. \& Sanz-Andres, A.} 2010 Energy harvesting
  from transverse galloping. {\em Journal of Sound and Vibration\/} {\bf
  329}~(14), 2873--2883.

\bibitem[Bearman(1984)]{Bearman1984}
{\sc Bearman, PW} 1984 Vortex shedding from oscillating bluff bodies. {\em
  Annual Review of Fluid Mechanics\/} {\bf 16}~(1), 195--222.

\bibitem[Beem {\em et~al.\/}(2013)Beem, Hildner \& Triantafyllou]{Beem2013}
{\sc Beem, Heather, Hildner, Matthew \& Triantafyllou, Michael} 2013
  Calibration and validation of a harbor seal whisker-inspired flow sensor.
  {\em Smart Materials and Structures\/} {\bf 22}~(1), 014012.

\bibitem[Bernitsas {\em et~al.\/}(2008)Bernitsas, Rasghavan, Ben-Simon \&
  Garcia]{Bernitsas2008}
{\sc Bernitsas, Michael~M, Rasghavan, Kamaldev, Ben-Simon, Y \& Garcia, EMH}
  2008 Vivace (vortex induced vibration aquatic clean energy): A new concept in
  generation of clean and renewable energy from fluid flow. {\em Journal of
  offshore mechanics and Arctic engineering\/} {\bf 130}~(4).

\bibitem[Connell \& Yue(2007)]{Connell2007}
{\sc Connell, B.S.H. \& Yue, D. K.~P.} 2007 Flapping dynamics of a flag in a
  uniform stream. {\em Journal of Fluid Mechanics\/} {\bf 581}~(-1), 33--67.

\bibitem[Dahl {\em et~al.\/}(2007)Dahl, Hover, Triantafyllou, Dong \&
  Karniadakis]{Dahl2007}
{\sc Dahl, JM, Hover, FS, Triantafyllou, MS, Dong, S \& Karniadakis, GE} 2007
  Resonant vibrations of bluff bodies cause multivortex shedding and high
  frequency forces. {\em Physical review letters\/} {\bf 99}~(14), 144503.

\bibitem[Holmes(1979)]{Holmes1979}
{\sc Holmes, Philip} 1979 A nonlinear oscillator with a strange attractor. {\em
  Philosophical Transactions of the Royal Society of London. Series A,
  Mathematical and Physical Sciences\/} {\bf 292}~(1394), 419--448.

\bibitem[Huang(1995)]{Huang1995}
{\sc Huang, L} 1995 Flutter of cantilevered plates in axial flow. {\em Journal
  of Fluids and Structures\/} {\bf 9}~(2), 127--147.

\bibitem[Lorenz(1963)]{Lorenz1963}
{\sc Lorenz, Edward~N.} 1963 Deterministic nonperiodic flow. {\em Journal of
  the Atmospheric Sciences\/} {\bf 20}~(2), 130--141.

\bibitem[Lugt(1980)]{Lugt1980}
{\sc Lugt, Hans~J} 1980 Autorotation of an elliptic cylinder about an axis
  perpendicular to the flow. {\em Journal of Fluid Mechanics\/} {\bf 99}~(04),
  817--840.

\bibitem[Lugt(1983)]{Lugt1983}
{\sc Lugt, Hans~J} 1983 Autorotation. {\em Annual Review of Fluid Mechanics\/}
  {\bf 15}~(1), 123--147.

\bibitem[Modarres-Sadeghi {\em et~al.\/}(2011)Modarres-Sadeghi, Chasparis,
  Triantafyllou, Tognarelli \& Beynet]{Modarres2011}
{\sc Modarres-Sadeghi, Y, Chasparis, F, Triantafyllou, MS, Tognarelli, M \&
  Beynet, P} 2011 Chaotic response is a generic feature of vortex-induced
  vibrations of flexible risers. {\em Journal of Sound and Vibration\/} {\bf
  330}~(11), 2565--2579.

\bibitem[Nakamura(1990)]{Nakamura1990}
{\sc Nakamura, Y} 1990 Recent research into bluff-body flutter. {\em Journal of
  Wind Engineering and Industrial Aerodynamics\/} {\bf 33}~(1), 1--10.

\bibitem[Obligado {\em et~al.\/}(2013)Obligado, Puy \& Bourgoin]{Obligado2013}
{\sc Obligado, M, Puy, M \& Bourgoin, M} 2013 Bi-stability of a pendular disk
  in laminar and turbulent flows. {\em Journal of Fluid Mechanics\/} {\bf 728},
  R2.

\bibitem[Robertson {\em et~al.\/}(2003)Robertson, Li, Sherwin \&
  Bearman]{Robertson2003}
{\sc Robertson, I, Li, L, Sherwin, SJ \& Bearman, PW} 2003 A numerical study of
  rotational and transverse galloping rectangular bodies. {\em Journal of
  fluids and structures\/} {\bf 17}~(5), 681--699.

\bibitem[Sipcic(1990)]{Sipcic1990}
{\sc Sipcic, Slobodan~R} 1990 The chaotic response of a fluttering panel: The
  influence of maneuvering. {\em Nonlinear Dynamics\/} {\bf 1}~(3), 243--264.

\bibitem[Spyrou \& Thompson(2000)]{Spyrou2000}
{\sc Spyrou, KJ \& Thompson, JMT} 2000 The nonlinear dynamics of ship motions:
  a field overview and some recent developments. {\em Philosophical
  Transactions of the Royal Society of London. Series A: Mathematical, Physical
  and Engineering Sciences\/} {\bf 358}~(1771), 1735--1760.

\bibitem[Townsend \& Shenoi(2013)]{Townsend2013}
{\sc Townsend, Nicholas~C \& Shenoi, RA} 2013 Modelling and analysis of a
  single gimbal gyroscopic energy harvester. {\em Nonlinear Dynamics\/} pp.
  1--16.

\bibitem[Van~Oudheusden(1995)]{Van1995}
{\sc Van~Oudheusden, BW} 1995 On the quasi-steady analysis of
  one-degree-of-freedom galloping with combined translational and rotational
  effects. {\em Nonlinear Dynamics\/} {\bf 8}~(4), 435--451.

\bibitem[Van~Oudheusden(1996)]{Oudheusden1996a}
{\sc Van~Oudheusden, BW} 1996 Rotational one-degree-of-freedom galloping in the
  presence of viscous and frictional damping. {\em Journal of fluids and
  structures\/} {\bf 10}~(7), 673--689.

\bibitem[Weymouth {\em et~al.\/}(2006)Weymouth, Dommermuth, Hendrickson \&
  Yue]{Weymouth2006}
{\sc Weymouth, Gabriel~D., Dommermuth, Douglas~G., Hendrickson, Kelli \& Yue,
  Dick K.-P.} 2006 Advancements in cartesian-grid methods for computational
  ship hydrodynamics. In {\em 26th Symposium on Naval Hydrodynamics\/}. Rome,
  Italy.

\bibitem[Weymouth \& Triantafyllou(2013)]{Weymouth2013JFM}
{\sc Weymouth, Gabriel~D. \& Triantafyllou, M.~S.} 2013 Ultra-fast escape of a
  deformable jet-propelled body. {\em Journal of Fluid Mechanics\/} {\bf 721},
  367--385.

\bibitem[Weymouth \& Yue(2011)]{Weymouth2011JCP}
{\sc Weymouth, Gabriel~D. \& Yue, Dick K.-P.} 2011 Boundary data immersion
  method for cartesian-grid simulations of fluid-body interaction problems.
  {\em Journal of Computational Physics\/} {\bf 230}~(16), 6233--6247.

\bibitem[Wibawa {\em et~al.\/}(2012)Wibawa, Steele, Dahl, Rival, Weymouth \&
  Triantafyllou]{Wibawa2012}
{\sc Wibawa, M.~S., Steele, S.~C., Dahl, J.~M., Rival, D.~E., Weymouth, G.~D.
  \& Triantafyllou, M.~S.} 2012 Global vorticity shedding for a vanishing wing.
  {\em Journal of Fluid Mechanics\/} {\bf 695}, 112--134.

\bibitem[Williamson \& Govardhan({2004})]{Williamson2004}
{\sc Williamson, CHK \& Govardhan, R} {2004} Vortex-induced vibrations. {\em
  Annual Review of Fluid Mechanics\/} {\bf {36}}, {413--455}.

\end{thebibliography}

\label{lastpage}
\end{document}